\documentclass{aa}
\usepackage[utf8]{inputenc}
\usepackage{slashbox}
\usepackage{natbib}
\usepackage[colorlinks]{hyperref}
\usepackage{xcolor}
\usepackage{ulem}
\usepackage{booktabs}
\hypersetup{
    citecolor=[RGB]{44, 117, 255},  
    linkcolor=[RGB]{231, 62, 1},
    urlcolor=[rgb]{0,0,1}
}

\usepackage[]{color}

\definecolor{blue}{RGB}{0, 165, 255}

\definecolor{green}{RGB}{106,180,153}

\definecolor{violet}{RGB}{212,115,212}

\definecolor{orange}{RGB}{255,127,0}

\definecolor{red}{RGB}{255,0,0}

\newcommand{\plk}{\textit{Planck }}

\makeatletter
\renewcommand*\aa@pageof{, page \thepage{} of \pageref*{LastPage}}
\makeatother

\title{Questioning Planck-selected star-forming high-redshift galaxy protoclusters and their fate}

\titlerunning{}
\author{
  C.~Gouin\inst{1,2}\thanks{E-mail:~\tt{celinegouin@kias.re.kr}},  
  N.~Aghanim\inst{2},  
  H.~Dole\inst{2}, 
  M.~Polletta \inst{3}, 
  C.~Park  \inst{1}
}
\institute{
School of Physics, Korea Institute for Advanced Study (KIAS), 85, Hoegiro, Dongdaemun-gu, Seoul, 02455, Republic of Korea
\label{inst1}
\and
Université Paris-Saclay, CNRS, Institut d’Astrophysique Spatiale, 91405, Orsay, France
\label{inst2}
\and
\label{inst3}
INAF - Istituto di Astrofisica Spaziale e Fisica cosmica (IASF) Milano, via A. Corti 12, 20133 Milan, Italy
}

\date{\today}

\abstract
{
About 2100 star-forming galaxy protocluster candidates at $z{\sim}$1--4 were identified at submillimeter (sub-mm) wavelengths in the \plk all sky survey.
Follow-up spectroscopic observations of a few candidates have confirmed the presence of actual galaxy overdensities with large star-formation rates.

In this work, we use state-of-the-art hydrodynamical simulations to investigate whether the \plk high-$z$ sub-mm sources (PHz) are progenitors of massive clusters at $z=0$.
To match the PHz sources with simulated halos, we select the most star-forming (SF) halos in 19 redshifts bins from $z=3$ to $z=1.3$ in the TNG300 simulation of the IllustrisTNG project.
At each redshift, the total star formation rate (SFR) of the simulated protocluster candidates is computed from the SFR of all the galaxies within an aperture corresponding to the \plk beam size, including those along the line-of-sight. 

The simulations reproduce the \plk derived SFRs as the sum of both, the SFR of at least one of the most SF high-$z$ halo, and the average contribution from SF sources along the line-of-sight.
Focusing on the spectroscopically confirmed $z \sim 2$ PHz protoclusters, we compare the observed properties of their galaxy members with those in the most SF simulated halos. We find a good agreement in the stellar mass and SFR distributions, and in the galaxy number counts, but the SFR-stellar mass relation of the simulated galaxies tends to be shifted to lower SFRs with respect to the observed one.  
Based on the estimated final masses of the simulated halos, we infer that between 63\% and 72\% of the \plk selected protoclusters will evolve into massive galaxy clusters by $z=0$. Despite contamination from star-forming galaxies along the line of sight, we thus confirm the efficiency of \plk to select star-forming protoclusters at Cosmic Noon with the simulations, and provide a new criterion for selecting the most massive cluster progenitors at high-$z$, using observables like the number of galaxy members and their SFR distribution.
}


\authorrunning{Gouin et al.}

\begin{document}
\maketitle

\section{Introduction}

Observing the early stages of the Universe is of prime importance for testing models of large-scale structure formation and evolution.
Clusters of galaxies are the most massive virialized structures at the present time. Their progenitors, often referred to as protoclusters, are not yet virialized and spatially extended structures given that they are merging and collapsing in accordance with the hierarchical structure formation scenario \citep{Sheth1999}. 
Compared to galaxy clusters at $z=0$, which are mainly populated by red massive galaxies, their high-$z$ progenitors are supposed to host the peak of star-formation activity in the history of the Universe, by being responsible for more than 20\% of the cosmic star formation at $z > 2$ \citep[e.g.][]{Kauffmann2004,Madau2014,Chiang2017}. These high-$z$ star-forming (SF) environments are thus also key places for probing our current understanding of star formation and galaxy growth at the Cosmic Noon epoch ($1 < z < 3$) \citep{Chiang2017}. Probing galaxy clusters at their primordial evolutionary stage is therefore essential to understand the assembly history of clusters \citep[see e.g.][]{Cohn2008,Rennehan2020} and to open a window on this early stage of intensive SF activity \citep[][for a review]{Overzier2016}.

In the past decade, large efforts have been put in finding cluster progenitors, and thousands of candidates have been found, but their confirmation as protoclusters remains challenging. Today, only tens of protocluster candidates have been confirmed. They are typically drawn from a variety of selections, and span a wide redshift range, from $z{\sim}$1.5 to $>$6. Samples of homogeneously-selected and confirmed protoclusters over a wide redshift range are still missing, hampering statistical studies, and limiting our understanding of their evolutionary properties.
Low-redshift clusters are usually detected via their X-ray emission \citep[see e.g.][]{Rosati1998,Bohringer2004} or through the Sunyaev-Zel'dovich (SZ) effect \citep[see e.g.][]{Planck2014_cluster,SPT_2015} from their hot plasma \citep{Rosati2002}, or in the optical band through overdensities of red galaxies \citep[see e.g.][]{Redmapper}. 
In contrast, protoclusters do not have a sufficiently massive and hot plasma in their core, making both their X-ray and SZ effect signals below the sensitivity reached by current instruments, and common cluster detection methods inefficient.
Different approaches have been proposed over the past decade to find protocluster, such as narrow-band imaging to detect H$\alpha$ or Ly$\alpha$ emission from SF galaxies at a specific redshift \citep[e.g.][]{Daddi2021,Shi2021,Zheng2021}, extended overdensities of star-forming galaxies \citep[e.g.][]{Chiang2014,McConachie2022}, and redshift searches around high-redshift radio galaxies (HzRGs) radio or sub-millimetre galaxies (SMGs)~\citep[e.g.]{Umehata2015,Oteo2018,An2021,Kalita2021}.
Nevertheless, these protocluster searches remain limited to a small fraction of the sky, such as the COSMOS field \citep{Ata2021} and the \textit{Hubble Ultra-Deep Field}, strongly biased by their selection method, and inhomogeneous in terms of available data, making it difficult to build a homogenous sample and to carry out meaningful comparisons across multiple protoclusters.

In the last two decades, more and more protocluster candidates highly emitting at rest-frame far-infrared (FIR) wavelengths, corresponding to observed sub-millimeter (sub-mm) frequencies, have been discovered \citep[e.g.][]{Lagache2005,Beelen2008,Ivison2013,Vieira2010,Smail2014,Dowell2014,Dannerbauer2014,Umehata2014,Hill2020,Rotermund2021}.  
Given the expected high star-formation activity in high-$z$ dense environments at the Cosmic Noon epoch, the sub-mm/mm band is indeed an ideal window for probing high-$z$ dusty SF galaxies (DSFGs) as supported by physical modeling of galaxy evolution \citep[e.g.][]{Negrello2005}.
These high-$z$ IR-luminous galaxies are, as expected, highly star-forming with star-formation rates (SFRs) of up to several thousands of $M_{\odot} /yr$.

In this context, the sub-mm measurements achieved by the \plk mission offer a unique opportunity to statistically analyse the most luminous sub-millimeter sources at high redshift over a large sky fraction.
By using high frequency (between 353 and 857 GHz) maps from the \plk mission, \cite{Planck2016_proto} have selected 2151 bright sub-mm sources, so called PHz, over the cleanest $26\%$ of the sky. These \plk sources provide a homogeneously-selected sample of protocluster candidates at $z\sim 1-4$, and thus, constitute a powerful sample for studying the early stage of cluster formation at their peak of SF activity. Indeed, significant overdensities of DSFGs have been revealed by cross-matching \textit{Herschel} and \plk data \citep[see e.g.][]{Clements2014,Greenslade2018,Cheng2019,Lammers2022}. Such galaxies are 
thought to be the progenitors of the massive elliptical galaxies found in local clusters.

Nevertheless, the abundance (about $0.2 \ sources/deg^2$) and flux densities (with IR luminosity typically around few $10^{14}L_{\odot}$) of these \plk high-redshift source candidates is far larger than expected from $\Lambda$CDM models and structure formation scenarios~\citep{Negrello2017}. 
Even if a small fraction of PHz candidates is supposed to be strongly gravitationally lensed galaxies, the number of bright sub-mm sources is significantly higher than the number of protoclusters predicted by cosmological models of large-scale structure growth.
As explained by \cite{Negrello2017}, using an analytical formalism based on galaxy evolution models \cite{Negrello2005}, the expected count of sub-mm luminous protoclusters from standard $\Lambda$CDM model is far below the statistics derived from the \plk detections. 
\cite{Negrello2017} find that this discrepancy can be explained by a positive Poisson fluctuation of dusty high-$z$ sources within the \plk beam. 

In this work, we investigate whether both, the estimated SFRs of the PHz sources from \cite{Planck2016_proto,Planck2015herschel}, which are about ten times larger than common sub-millimeter detections (around few $10^4 \ M_{\odot} /yr$), and the PHz follow-up observations can be explained by hydrodynamical simulation.
More specifically, we will investigate the possibility that the PHz sources are the result of chance projection of multiple SFGs along the line of sight.
In details, we will see if the \plk high-z sources and their follow-up observations can be explained by using the distribution of SF halos in state-of-the-art hydrodynamical simulation, and, in such a case, what is their expected evolution.
Indeed, the progress of hydrodynamical cosmological simulations in the recent years has opened a new window for interpreting protocluster galaxy observations, for example, \cite{Lim2021} compared seven structures from \cite{Casey2016}, and \cite{Araya2021} with the expected protocluster detections from the Hyper Suprime-Cam Subaru Strategic Program. For the first time, we explore possible interpretations of the \plk high-$z$ SF protocluster candidates via their integrated SFRs, galaxy member properties as derived from spectroscopic follow-up studies, and fate at $z=0$, by using  hydrodynamical simulations.

The paper is organized as follows.
In Sect. 2, we describe the \plk selection of high-$z$ sub-mm sources and their follow-up observations. 
In Sect. 3, we present our selection of the most SF high-$z$ objects from the TNG300 simulation as our simulated protocluster candidate sample. We detail a parametric toy model to compute their SFR by taking into account angular aperture size, and the foreground and background contamination along a fiducial line-of-sight (LOS). 
In Sect. 4, we compare the \plk protocluster candidates to our high-$z$ SF halo sample, by investigating both the integrated SFR within the \plk beam, and their galaxy member properties as derived from PHz follow-up observations. We also explore the fate of \plk protocluster candidates, by probing the evolution of our simulated sample up to $z=0$.
In Sect. 5, we discuss the potential limitations and biases of \plk source detection and follow-up observations. We also compare our finding with recent works probing star formation of protoclusters in simulation.
In Sect 6. , we summarize our key results on probing the \plk protocluster candidates in simulations.

\section{Observational data sets \label{SEC:DATA}}

In this section, we introduce the 2151 \plk high-$z$ (PHz) SF protocluster candidates, and the follow-up observations of three of them.

\subsection{The \plk selection of protocluster candidates}

\cite{Planck2016_proto} have searched for bright sub-mm sources, with colors consistent with $z{\sim}$1--4 using the high frequency all-sky maps obtained from the \plk mission with an angular resolution of about 5 to 10 arcmin, and over the cleanest 26\% of the extragalactic sky. 
Typical sub-mm SEDs of high-$z$ ($1<z<4$) sources are expected to peak from 353 to 857 GHz \citep[depending on their redshift, see figure 2 of][]{Planck2016_proto}, making the \plk high frequency coverage optimal for statistical detections of dusty FIR luminous sources at high redshift. 
Nevertheless, given that these sources are embedded in Galactic cirrus, cosmic infrared background (CIB), and cosmic microwave background (CMB), \cite{Planck2016_proto} have developed a dedicated approach to remove the CMB component, the Galactic and low-$z$ CIB component, and to optimize the detection of a signal in excess at 545 GHz. 
Whereas the CMB cleaning procedure removes also sub-mm sources at very high-$z$ ($z>4$), the Galactic Cirrus cleaning strongly reduces the contamination of sub-mm sources at low-$z$ ($z<1$). Following the cleaning procedure, cleaned maps at 857, 545, 353 and 217 GHz are obtained. An excess map at 545\,GHz is then produced by subtracting the linear interpolation of the two surrounding bands (353 GHz and 857 GHz) from the 545\,GHz maps.
The PHz sample is finally constructed in a systematic way, by requiring a simultaneous detection in the 545 GHz excess map (with $SNR>5$), and in the 857, 545, and 353 GHz cleaned maps (with $SNR > 3$), and absence of signal at 100 GHz (with $SNR < 3$).
The PH$z$ is thus color-selected (and not flux-limited) and selects colors compatible with galaxies spectral energy distributions at redshifts $1<z<3$ \citep{Planck2016_proto}. 

The PHz sources are unresolved sub-mm peaks with an angular resolution between 5 and 10 arcmin (depending on the frequency) detected in four \plk wavelength bands. The flux density is computed for each detected sources via aperture photometry in the four cleaned maps (at 857, 545, 353 and 217 GHz). Notice that, given that flux densities are computed in cleaned maps (and not in the excess map due to calibration difficulties), they include the signal emitted along the line of sight from $z{>}1$ to ${<}$4 sources.
They might also be affected by attenuation and contamination from systematics effects as discussed in \cite{Planck2016_proto}.

For each PHz source, several photometric redshifts are estimated by fitting the sub-mm spectral energy distribution (SED) with a modified blackbody model at various dust temperatures.
In the present work, we use the results estimated by assuming a dust temperature of $30K$, in agreement with recent spectroscopic observations of high-z sub-millimeter galaxies \citep[see e.g.][]{Magnelli+14}. 
The redshift distribution of the \plk sources range from $z=1.3$ to $3$ for 90\% of the sample, and peaks at $z\sim 2$. 
Therefore, the PHz sample represents a precious resource to investigate the star-formation peak activity in the Universe \citep{Chiang2017}, and the early phases of cluster formation.

\begin{figure}
    \centering
    \includegraphics[width=0.5\textwidth]{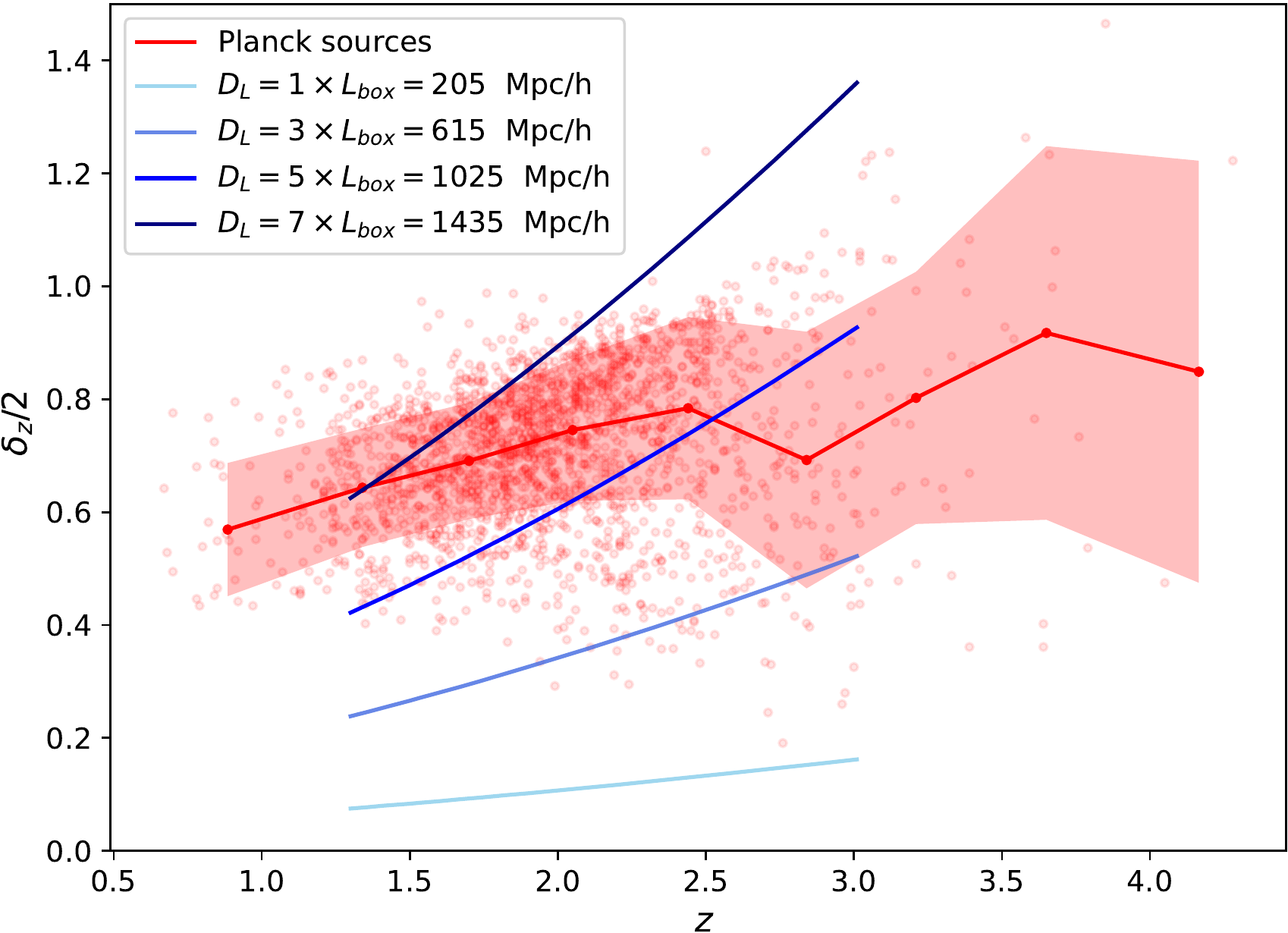}
     \caption{Redshift uncertainty, $\delta_z$/2, defined by the mean 1$\sigma$ uncertainty, as a function of redshift for all \plk sources (red circles). The average $\delta_z$/2 per redshift bin and its variance are shown as a red solid line, and light red filled area. 
     The redshift interval $\delta_z$/2 corresponding to a comoving distance of $D_{L}=$ 205, 615, 1025, and 1435 Mpc/h, between $z$\,=\,1.3 and 3.0, are shown as blue lines.
     \label{fig:PLANCK_deltaz_z}}
\end{figure}

The average uncertainty associated with the \plk photometric redshift estimates is about $\delta_z \sim 1.4$ and slowly increasing with the redshift, as illustrated in Fig. \ref{fig:PLANCK_deltaz_z}.  We define this \plk redshift uncertainty $\delta_z$ as the interval between the best-fit $z{-}1{\sigma}$ and $z{+}1{\sigma}$, obtained during the SED fitting procedure as presented in \cite{Planck2016_proto}.
The derived photometric redshift uncertainty reflects the approximation introduced by the assumption that the measured flux densities are produced by a single source at a specific redshift, rather than by multiple sources along the LOS.
Following this assumption, we highlight in Fig. \ref{fig:PLANCK_deltaz_z} the redshift uncertainty which correspond to different integral of comoving distance along the LOS in blue lines ($D_L=$ 205, 615,1025, and 1435 Mpc/h). In general, a integration of a comoving distance of 1025 Mpc/h along the LOS is the good choice to reproduce the average redshift uncertainty of Planck sources (i.e.; $\delta_z \sim 1.4$). However, the redshift uncertainties of Planck sources at redshift $z<2$ appear to be better represented by comoving distances between 1025 and 1435 Mpc/h. 

\begin{figure}
    \centering
    \includegraphics[width=0.485\textwidth]{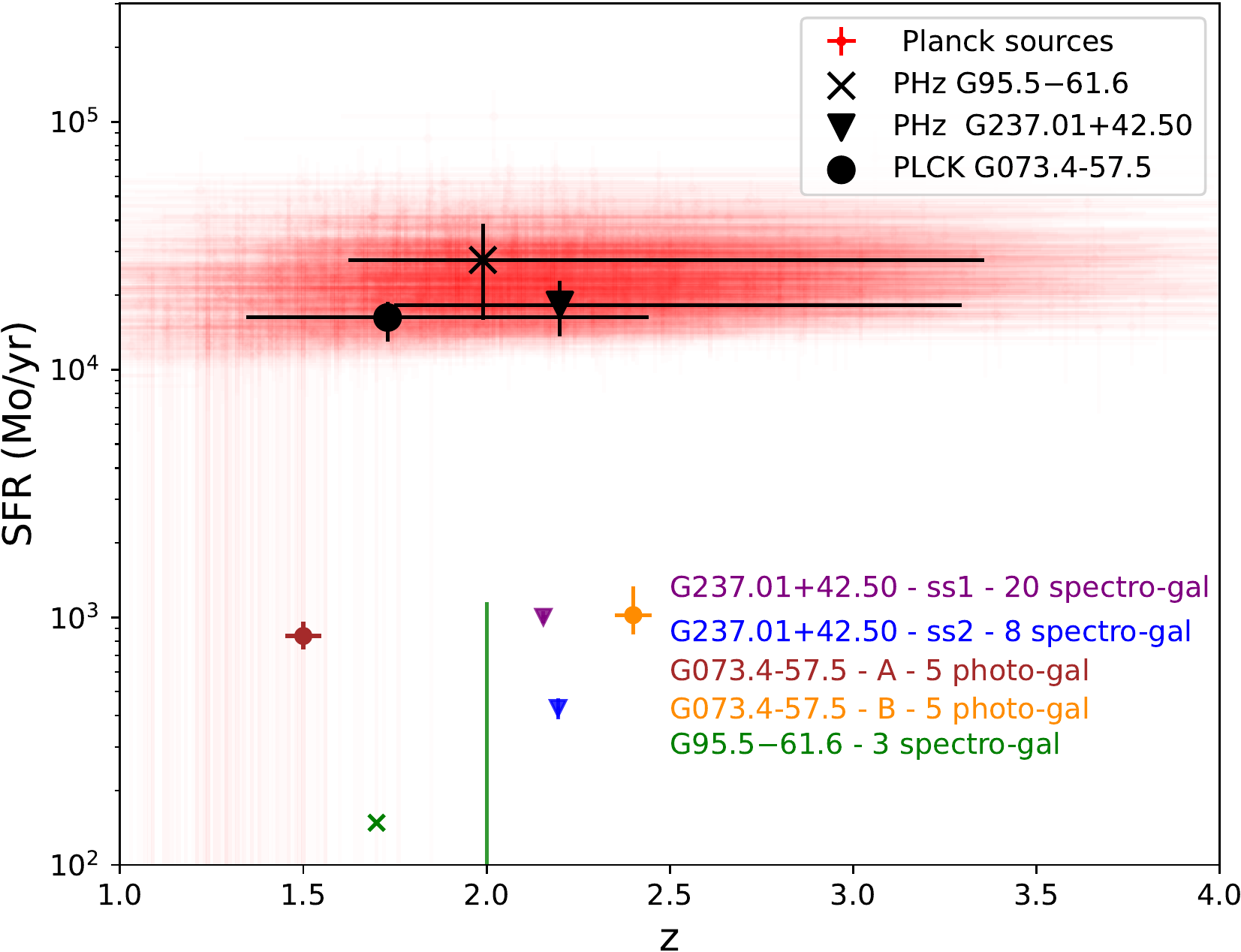}
     \caption{Star formation rates as a function of redshift, and associated uncertainties, for the $2151$ high-$z$ SF Planck sources (red crosses). The \plk estimates for three sources with spectroscopically confirmed structures are highlighted with large symbols: G237.01+42.50 (upside-down black triangle), G95.5-61.6 (black cross), and G073.4-57.5 (black circle). The values derived from the spectroscopically confirmed members of these three sources are shown with colored symbols as noted. 
     \label{fig:DATA_FOLLOWUP}}
\end{figure}

The PHz catalog \cite{Planck2016_proto} includes also the FIR luminosity associated with each source derived by integrating the best-fit modified blackbody model between 8 and 1000$\mu$m, and the star-formation rate (SFR) derived following the prescription from \cite{Kennicutt1998}. 
In Fig. \ref{fig:DATA_FOLLOWUP}, we show the total SFR and associated uncertainties as a function redshift for the entire PHz sample. 
The estimated SFRs are extremely high, with the 16$^{th}$ and 84$^{th}$ percentile being 1.6$\times$10$^4$\,M$_{\odot}$\,yr$^{-1}$, and 3.2$\times$10$^4$\,M$_{\odot}$\,yr$^{-1}$ (for a dust temperature equal to $30K$). 
As discussed in the introduction, we resort to large volume hydrodynamical simulation to investigate the possible origins of these high SFR.

\begin{table*}[]
    \centering
    \caption{Main properties of the structures found in the PHz sample}
    \begin{tabular}{l c c c c c c}
\hline\hline
        Name & $z$ range & size [$\textrm{arcmin}^2$] & $N_{gal}$ & $\Sigma$SFR [$M_{\odot}/yr$]& ${\Sigma}M_{*}$ [$10^{11} \ M_{\odot}/h$] \\
        \hline
        G237.01+42.50 \textit{ss1} & 2.15-2.164 & $9.5 \times 9.3$ & 20 & $1002\pm58$ & $7.3\pm0.5 $ \\
        G237.01+42.50 \textit{ss2} & 2.19-2.20 & $9.9 \times 10.6$ & 8 & $429\pm41$ & $2.1\pm0.2 $   \\
        G073.4-57.5 \textit{A} & $\sim1.5$ & $2.4 \times 2.4$ & 5 & $840^{+120}_{-100}$ & $5.8^{+1.7}_{-2.4} $ \\
        G073.4-57.5 \textit{B} & $\sim2.4$ &  $2.4 \times 2.4$ & 5 & $1020^{+310}_{-170}$ & $4.2^{+1.5}_{-2.1} $  \\
        G95.5-61.6  & $1.677-1.684$ & $1 \times 1$ & 3 & $148\pm10$ & $4.0$  \\
        \hline
    \end{tabular}
    \tablefoot{The properties of the structures found through follow-up observations of three \plk high-$z$ protocluster candidates are from \cite{Polletta+21} (G237.01+42.50), \cite{Kneissl+19} (G073.4-57.5) and \cite{Flores-Cacho+16} (G95.5-61.6).}
    \label{tab:DATA}
\end{table*}

\subsection{PHz follow-up observations}

With the aim to investigate the nature of these \plk sources, follow-up observations with the Herschel Space Observatory and with the Spitzer Space Telescope of, respectively, 228, and 92 PHz sources have been carried out~\citep{Planck2015herschel,Martinache2018}. In addition, for a handful of PHz sources optical and NIR imaging and spectroscopic observations have been obtained. At the present time, only three PHz sources have been spectroscopically confirmed. 

First, PHz G95.5-61.6 was observed by \cite{Flores-Cacho+16} from optical to sub-millimeter wavelengths, and with targeted optical-NIR spectroscopic observations that have revealed an overdensity of sub-mm sources associated with two clumps of galaxies at high-redshift.
They found one structure at $z\sim 1.7$ with three spectroscopically confirmed galaxy members, and a second one at $z\sim 2.0$ with six confirmed members. Unfortunately the SFR of the structure at $z\sim 2.0$ has not been estimated given that two of the six galaxies are fully blended in their aperture photometry. The SFR, stellar mass, and redshift obtained from each member galaxies of the structure at $z\sim 1.7$ are presented in Table 3 of \cite{Flores-Cacho+16}, and will be used in this analysis. 

Secondly, the \plk source PHz G073.4-57.5 has been observed with ALMA and fully investigated by \cite{Kneissl+19}. They found that this sub-mm emitted source is at least composed of two distinct SF structures (called $A$ and $B$) along the line-of-sight (LOS), with five galaxy members for each: one at $z \sim 1.5$ and the second one at $z \sim 2.4$. Beyond these two SF high-$z$ groups, they also detect eight luminous SF galaxies along the LOS.
Notice that this follow-up observation is not spectroscopic, but the photometric redshift uncertainty is about $0.15$. They also provided  star-formation rates, and stellar masses for each galaxy in Table 5 of \cite{Kneissl+19}. These measurements will be compared here with those of the simulated protocluster members. 

More recently, a third PHz source, PHz G237.01+42.50 has been deeply investigated by \cite{Polletta+21}. This source is located in the COSMOS field and contains at least 31 spectroscopically confirmed galaxies at redshift around $z\sim2.16$.
In details, this source appears to be the sum of two substructures or protocluster regions: one clump of 20 galaxy members at $2.15<z<2.165$, and the second one with 8 galaxies at $2.19<z<2.20$.
Galaxies of these two structures (called $ss1$ and $ss2$) are mostly blue SF galaxies with SFR and stellar masses consistent with the main sequence, and with a significant fraction ($20\pm10\%$) of Active galactic nuclei (AGN). Galaxy member SFR and stellar masses used in this study are reported in Table 8 of \cite{Polletta+21}.

The total SFR, and stellar mass, and the redshift range of these three PHz sources containing five SF galaxy structures at high-$z$ are reported in Table \ref{tab:DATA}. 
In Fig. \ref{fig:DATA_FOLLOWUP}, we show the \plk measurements of these sources with black symbols.
Their associated SFRs and redshifts, as measured from the confirmed members in the follow-up observations, are instead shown with color symbols.
The follow-up observations indicate that some PHz sources contain at least two different structures aligned along the LOS. The follow-up observations also prove that the total SFR obtained by considering the identified SF galaxies in those structures is much lower than the SFR estimated from the \plk data. 

\section{Methods \label{SEC:METHOD}}

To understand the nature of the \plk sources, and reconcile their properties with those derived from the follow-up observations, we resort to large volume cosmological simulations \citep[see e.g.][]{Granato2015,DIANOGA2020}.
Since not all PHz sources might contain a high-$z$ protocluster, indeed a small percentage of them are strongly lensed galaxies~\citep{Planck2015herschel,Canameras2015,Canameras2021}, in the simulations we will question the \plk selection without imposing a priori that the simulated object will evolve into a massive cluster by $z=0$. The fate of the simulated structures will be investigated and discussed later. The issue of how to select true protoclusters in simulations has been previously discussed by \cite{Lim2021} using the TNG300 simulations.
They concluded that simulated objects selected by their total SFR, rather than by halo mass, and independently of their final mass at $z=0$, reproduce better protocluster observations. This is especially true for protoclusters discovered as high SF galaxy overdensities or because of their bright IR luminosity.

Starting from this point, we construct a simulated protocluster sample making use of the TNG300-1 simulation, selected on the their total SFR. For each simulated object, we compute the total SFR from the distribution of SF galaxies within different volumes, as presented below.

\subsection{TNG300-1 simulation}

The IllustrisTNG project is a series cosmological hydrodynamical simulations that simulate the formation and evolution of cosmic structures from high redshift to the present time \citep{ILLUSTRIS_TNG}. The numerical models that govern the key physical processes relevant for galaxy formation and evolution are described in \cite{Pillepich2018}. We refer to them for details on the SFR computation per gas cells in the simulations of IllustrisTNG project. 
We focus here on TNG300-1 simulation of the IllustrisTNG project (TNG300-1 hereinafter), for which the comoving size of a simulation box is about $300$ Mpc, and the resolution in mass is about $m_{\text{DM}} = 4.0 \times 10^7 M_{\odot}/h$. 
The TNG300-1 is the largest simulation box of the IllustrisTNG series with the best spatial and mass resolution. This choice is optimal for probing SF structures at high redshift through their galaxy distribution. Indeed, as discussed by \cite{Lim2021}, the computation of SFR can be affected by the spatial resolution of hydrodynamical simulations.
The TNG300-1 simulation has 100 available snapshots, each consisting of one simulated box extracted at a specific time step from $z\sim 20$ to $z=0$. 
For each snapshot, the TNG300-1 provides a halo catalogue, with halos identified by the friends-of-friends (FoF) algorithm \citep{Davis1985}, and subhalo catalogues derived from the Subfind algorithm \citep{subfind}. 
We consider here galaxies all subhalos with a stellar mass and a SFR larger than $0$, and halos as objects detected by the FoF algorithm. 
Halos are constituted of both a main subhalo (the most massive subgroup of each halo) and other subhalos associated with it by the FoF algorithm.
The simulation assumes a cosmology consistent with results from \cite{Planck2016}, such that $\Omega_{\Lambda,0} = 0.6911$, $\Omega_{m,0} = 0.3089$, $\Omega_{b,0} = 0.0486$, $\sigma_{8} = 0.8159$, $n_s = 0.9667$ and $h = 0.6774$.

\subsection{Simulated most SF protocluster candidates in TNG300-1}

To question the \plk selection in the simulations, we consider the most SF objects at various redshifts, even if they might not evolve into massive halos by the present time. 
The SFR selection is performed on the SFR of the halo, referred later as $SFR_{FoF}$. 
This SFR value, provided by the TNG300 simulation, is the sum of the individual SFRs of all gas cells in a FoF group. 
We then construct a sample of high-$z$ SF objects, by selecting the $30$ most SF halos from $z=1.3$ to $z=3$ in IllustrisTNG.
This redshift interval is chosen to match the redshift range covered by \plk sources. 
Indeed, more than 90\% of the PHz have redshifts 1.3${<}z{<}$3. This redshift range is covered by $19$ snapshots in the simulation.
The choice of $30$ SF halos at each redshift is based on the number of massive clusters at $z=0$ in TNG300-1. Indeed, there are approximately $30$ galaxy clusters with mass\footnote{$M200$ is defined as the total mass of a group enclosed in a sphere whose mean density is 200 times the critical density of the Universe, at the time the halo is considered.} $M_{200}>2.5 \times 10^{14} M_{\odot}/h)$ \citep[similar selection as ][which have used TNG300]{Lim2021}.
Therefore, our simulated sample of \textit{"mock protoclusters candidates"} from TNG300-1 is constituted of the $570$ most SF objects at 1.3${<}z{<}$3.

\subsection{The fate of the simulated most SF objects}

Starting from our simulated sample of the 570 most SF objects from $z=1.3$ to $z=3$, we aim to probe their fate at $z=0$. In order to test if they are actual protoclusters, e.g. progenitors of massive galaxy clusters at $z=0$, we focus on their descendant at the last snapshot of the simulation ($z=0$).

To derive the final mass of the descendant of each simulated object, we use the merger tree of the associated subhalos computed using the SubLink algorithm, as provided by IllustrisTNG \citep[see][for details on merger tree computation]{Rodriguez2015}.  
For each high-$z$ SF halo in our simulated sample, we extract the merger tree of its main subhalo. Following the main principal branch of each merger tree, we can find its descendant at $z=0$. 
To do so, we consider the last descendant at $z=0$ in each merger tree.
From the tree identification number, we can identify the last descendant subhalo at $z=0$, and thus, the associated halo at $z=0$. 
Through this procedure, for each halo in our high-$z$ SF halo sample, we can derive the properties of the last descendant halo such as the mass $M_{200}(z=0)$. 

Given the fact that we are using a subhalo merger tree, we can also determine whether a subhalo will become the main subhalo or a satellite substructure inside a given halo at $z=0$. 
Indeed, a main subhalo at a given redshift will not necessary remain the main subhalo during its whole merging history. Given that halos (and subhalos) are merging during cosmic time, a main subhalo at high redshift can become the substructure of another halo at $z=0$ (by being accreted or by a merger event).
In Sect \ref{SEC:RESULTS}, we will thus investigate the fate of our 570 SF objects by considering with which main halo they will be associated at $z=0$: either a main subhalo or a substructure; and their final halo mass: either a cluster-mass ($M_{200c}(z=0)>10^{14} M_{\odot}$), a group-mass ($M_{200c}(z=0)=10^{14-13} M_{\odot}$) or a low-mass object ($M_{200c}(z=0)<10^{13} M_{\odot}$).

\subsection{Halo properties per volume}

\begin{figure*}
    \centering
    \includegraphics[height=47mm]{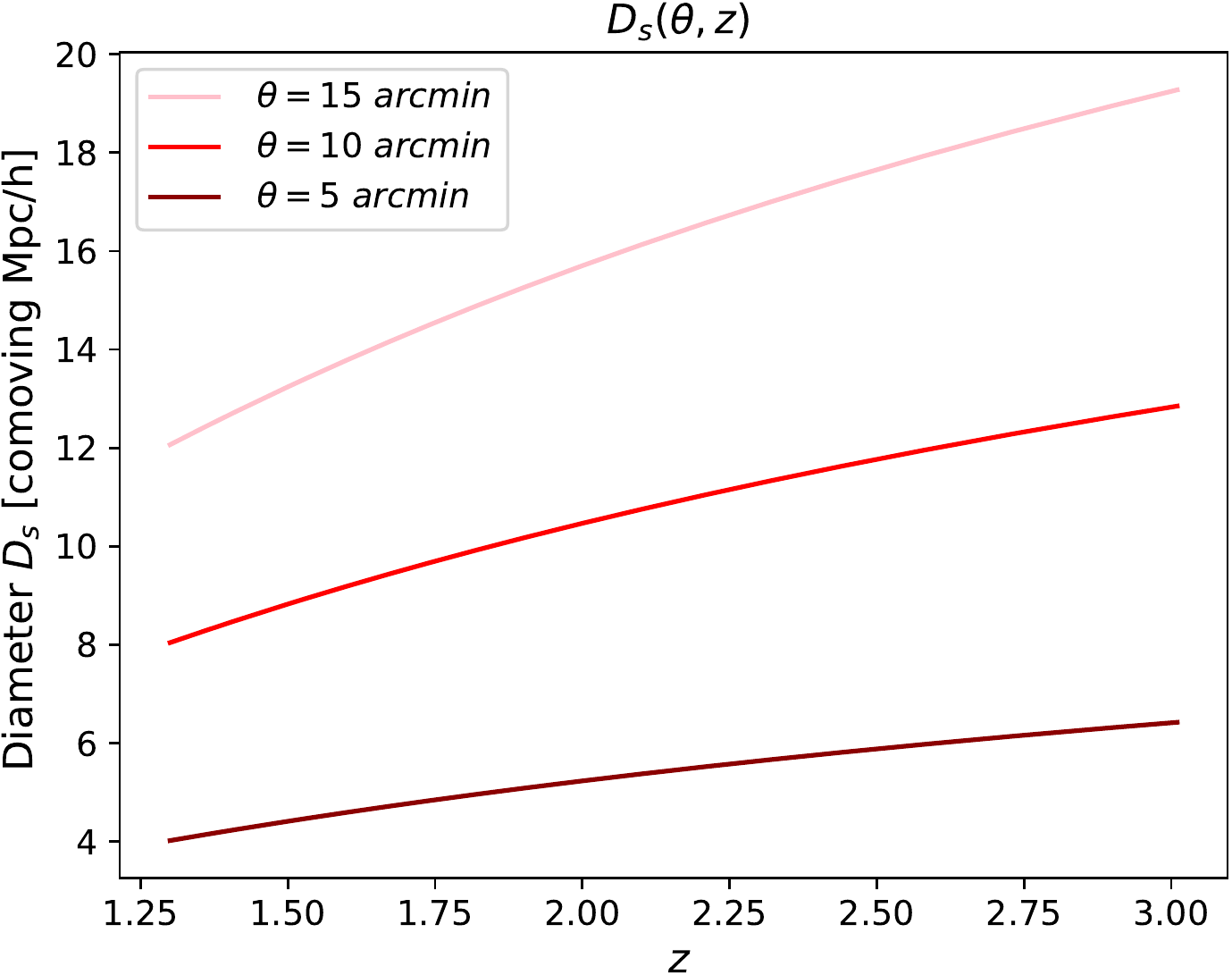}
    \includegraphics[height=47mm]{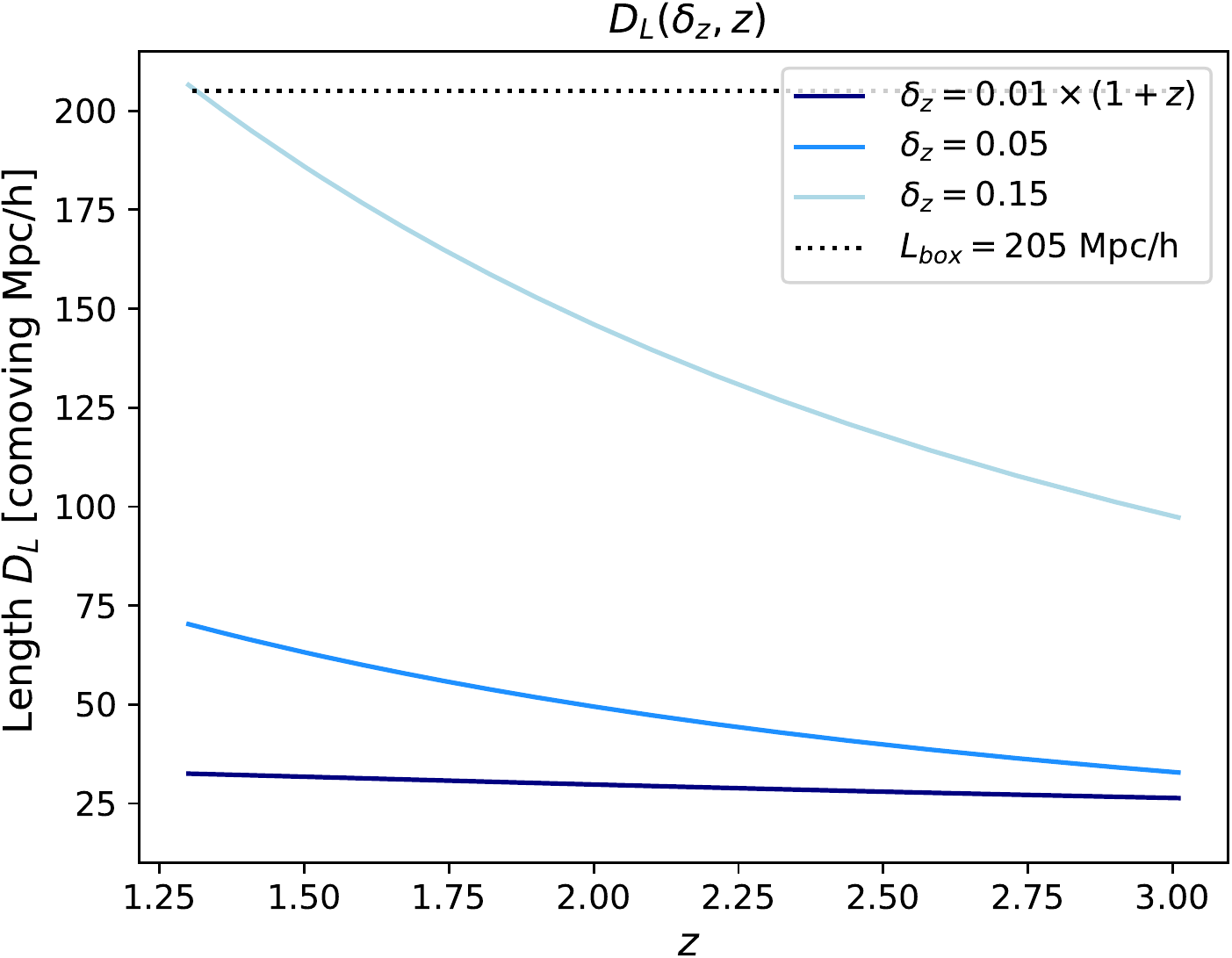}
    \includegraphics[height=47mm]{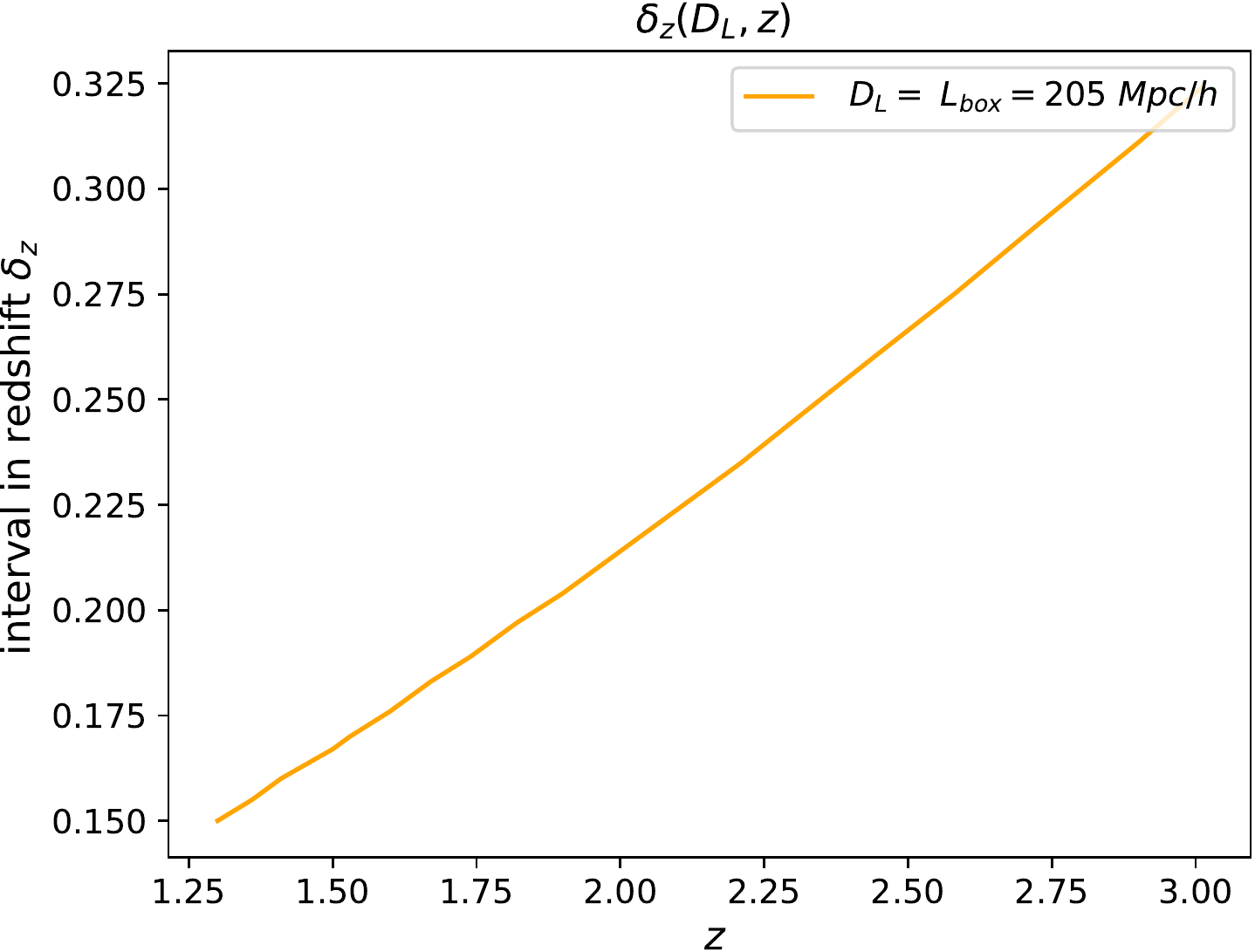}
     \caption{Parameters of the cylinder parametric model, and of the simulated box as a function of redshift.
     \textit{Left panel:} Cylinder diameter $D_s$ in comoving coordinates for three different angular size $\theta=5,10$, and $15$ arcmin. 
     \textit{Middle panel:} Cylinder length as comoving distance between two redshifts $z-\delta_z/2$ and $z+\delta_z/2$ for three different values of $\delta_z$ as noted. The length of the simulation box is shown with a dotted black line.
     \textit{Right panel:} Redshift interval $\delta_z$ corresponding to the size of one simulation box $L_{box}= 205 \ $Mpc/h.
     \label{fig:Distance}}
\end{figure*}

Given the large number of observational effects (aperture size, LOS contamination) in protocluster detection, and potential systematics in the measurements of their properties  (flux limits, SED fitting, galaxy selection), it is extremely difficult to perform a fair and reliable comparison between protocluster observations and simulations. 
In order to estimate the total SFR of the simulated protoclusters, different methods have been considered. We can measure the total SFR within a sphere of physical radius $R=5 \ R_{500}$ \citep{Lim2021}, or with a fixed radius of $10 \ Mpc$ in comoving scale \citep{Yajima2022}, or with different physical radii of $1 \ Mpc$ and $100 kpc$ \citep{DIANOGA2020}, or either in boxes with physical size per side of $2 Mpc$ \citep{Granato2015}. This large diversity of SFR protocluster estimations in simulation is the result of both, different observational conditions, and the highly debated theoretical predictions of the extent of high-$z$ protoclusters  \citep[see e.g.][]{Chiang2017,Muldrew2015}.

In our case, we aim to compare the galaxy population of our high-$z$ simulated halo sample with three PHz sources for which follow-up observations have confirmed the presence of a structure (see Tab. \ref{tab:DATA}). 
Given the different aperture size and redshift range along the LOS, from both the measurements relative to the identified members and the \plk data, we develop a parametric model which integrates the galaxy distribution inside a cylinder around each of our most SF high-$z$ halo \citep[see also][for cylindrical parametric model to probe galaxy protocluster]{Lovell2018}. We develop this technique to be flexible in the computation of the total SFR of our simulated sample, and to be able to adapt our parametrisation of the cylinder to each observed PHz and to the different methodologies adopted in measuring their integrated properties.  

In practice for each SF halo of our sample, we select all galaxies inside a cylinder with diameter $D_s$ and length $D_L$, and centered on the center of mass of the halo. 
This cylindrical selection of galaxy distribution takes into account three basic parameters:
the aperture angle, the integration along the LOS, and a minimum SFR threshold applied to the galaxies.
We describe these different parameters in more detail below:
\begin{itemize}
    \item[$\bullet$] \textbf{Minimum galaxy SFR value $SFR_{min}$:}\\ 
    through this parameter only the galaxies with a SFR above such a threshold are taken into account to reflect the minimum SFR measured in the identified members by the follow-up observations. 
    \item[$\bullet$] \textbf{Angular aperture $\theta$:}\\ 
    The diameter of the cylinder $D_s$ is computed as the comoving transverse distance at a given redshift $z$ for an angular aperture of $\theta$. The redshift evolution of such a size is illustrated in the left panel of Fig. \ref{fig:Distance}, by considering different angular apertures $\theta=5,10$ and $15$ arcmin.
    \item[$\bullet$] \textbf{Integration along the LOS $D_L$:}\\
    By selecting the galaxies inside a cylinder of length $D_L$, we can artificially reproduce the integration along the LOS. This length can be converted into a redshift range $\delta_z$, that is the comoving distance along the LOS between two different redshifts: $z-\delta_z/2$ and $z+\delta_z/2$. The middle panel of Fig. \ref{fig:Distance} shows the value of $D_L$ as a function of redshift for three different values of $\delta_z$. 
    It is important to point out that there is maximum length of each cylinder given by the comoving size of a simulation box that is $D_L \leq L_{box} =205$ Mpc/h. 
    For each object, to compute the mean of each integrated quantity and an associated error, we consider three cylinders (with aperture $\theta$, and length $D_L$) centered on the halo center and oriented in three different directions (along the $x$-,$y$-, and $z$-axis) that mimic three fiducial LOS.
\end{itemize}
The total SFR of a given halo is thus simply the sum of the SFR of all galaxies inside a cylinder with parameters $\theta$, and $D_{L}$ and centered on the halo, such as:
\begin{equation}
 SFR  (\theta,D_{L})  = \sum^{N_{gal}}_i SFR_i (SFR_i>SFR_{min})  \,,
\end{equation}
with $N_{gal}$ the number of $i$ galaxies with $SFR_{i}>SFR_{min}$ contained inside a cylinder defined by the $\theta$, and $D_{L}$ parameters and centered on a given object at its snapshot. 
Similarly, the total stellar mass of a given halo is computed as:
\begin{equation}
 M_{\ast}  (\theta,D_{L})  = \sum^{N_{gal}}_i M_{\ast,i} (SFR_i>SFR_{min})  \,,
\end{equation}
with $M_{*,i}$ being the stellar mass of $i$ galaxies inside the same cylinder as above and centered on a given object at its snapshot.
The total stellar mass and SFR of each simulated SF halos are computed in cylinders defined by the parameters $\theta$, $\delta_z$, $SFR_{min}$ that reproduce at best the observational conditions of the PHz follow-up observations.

\subsection{Mimicking the \plk measurements in the simulations}

\begin{figure}
    \centering
    \includegraphics[width=0.5\textwidth]{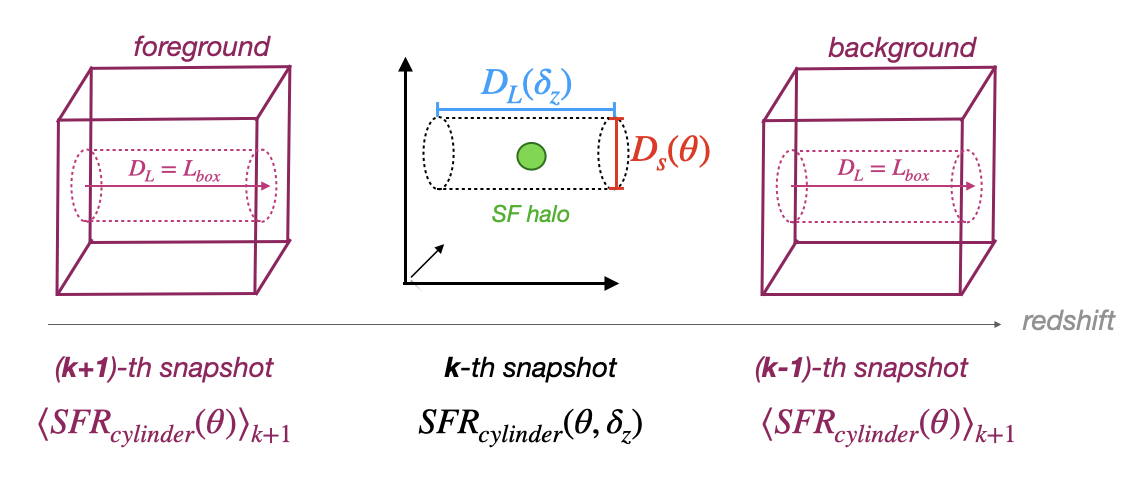}
     \caption{Illustration of the SFR computation method for a given SF halo in a $k$-th simulation snapshot at a given redshift $z_k$. 
     The total SFR of an object is computed by summing the SFR of the galaxies inside a cylinder of length $D_L(\delta_z)$ and diameter $D_s(\theta)$ centered on the halo. The contributions of $N$ background and foreground boxes along the LOS is considered by adding the mean total SFR computed inside cylinders with same parameters $D_L(\delta_z)$ and $\theta$ in the nearest snapshots at lower and higher redshifts $z_{k-1}$, and $z_{k+1}$.\label{fig:ILLUSTRATION_SFR}}
\end{figure}
To mimic the measurements carried out  with the \plk data on the full PHz sample~\citep{Planck2016_proto}, we consider cylinders with aperture $\theta$ consistent with the typical size of the PHz sources.
The \plk maps used to measure the sub-mm flux densities of the PHz sources have a resolution of $\sim$5 arcmin, and the sources' size have a major and minor axis FWHM that are ${\sim}10$ arcmin, and $5$ arcmin, respectively.
We, thus, choose an angular aperture $\theta=10$ arcmin. 
This angular size corresponds to cylinders with comoving diameters ranging from $8$ to $12$ Mpc/h at 1.3${<}z{<}$3 (see Fig. \ref{fig:Distance}). 
This size is in good agreement with the protocluster theoretical size predicted by \cite{Chiang2017}, given that it is approximately twice the radial distance where membership probability drops to 50\% in protoclusters at $1<z<3$.

The length of the cylinder that should be adopted to reproduce the \plk total SFRs of the PHz sources is defined by the estimated redshift uncertainty $\delta_z$ described in Sect.~\ref{SEC:DATA}. Such an uncertainty is, on average, equivalent to a comoving distance of $1025$ Mpc/h, as illustrated in Fig. \ref{fig:PLANCK_deltaz_z}. Such a distance is much longer than the maximum depth of a simulation box (i.e.; $L_{box} = 205 Mpc/h$). Indeed, for each snapshot of the simulation at redshift $z_{snap}$, the maximum redshift integral $\delta_z$ equivalent to a comoving distance of $205$ Mpc/h, is illustrated in the right panel of Fig. \ref{fig:Distance}.
Thus, to take into account the contribution from the galaxies along the LOS within $\delta_z$, we consider a cylinder length that combines five simulation boxes ($D_L = 5 L_{box}=1025$ Mpc/h), two in the foreground and two in the background of each simulated halo. 

For each snapshot $k$-th, we consider the total SFR from the most SF halo and that from the two foreground ($k+2$, $k+1$-th) and the two background ($k-1$, $k-2$-th) snapshots. The total SFRs from the foreground and background snapshots are obtained by averaging the total SFR given from all the galaxies within $100$ random cylinders, with parameters $\theta$\,=\,10 arcmin, and $D_L=L_{box}$, in those snapshot.
We illustrate these additional background and foreground contributions to our parametric computational method of halo SFR in Fig. \ref{fig:ILLUSTRATION_SFR}. The total SFR of a SF halo at $z_k$ is thus derived as:
\begin{multline}
    SFR (\theta,z_k) =  SFR (\theta,D_L=L_{box},z_k) \\
    + \sum_{j=k-2,k-1,k+1,k+2} \langle SFR (\theta,D_L=L_{box},z_{j})\rangle \,.
\end{multline}


\section{Results \label{SEC:RESULTS}}

We aim to compare our simulated sample of the 570 most SF halos from $z=1.3$ to $z=3$ and their associated properties, as computed by the cylinder parametric model described in Sect.~\ref{SEC:METHOD}, to both the \plk measurements of the full PHz sample, and to those obtained on three PHz from follow-up observations.

We first investigate the total SFR of the \plk protocluster candidates by including the contribution from foreground and background sources along LOS in the simulated sample.
Then, we compare the main properties (SFR, M$_{\ast}$, and number counts) of the galaxy members identified in the spectroscopically confirmed PHz protoclusters with those in the simulations. 
Finally, we analyze the evolution of the most SF simulated objects up to the present time. 
We also investigate which physical properties in the simulated halos can be used to better predict their evolution and the final halo mass at $z=0$.


\subsection{The effective SFR of the \plk SF protocluster candidates}

\begin{figure}
    \centering
    \includegraphics[width=0.49\textwidth]{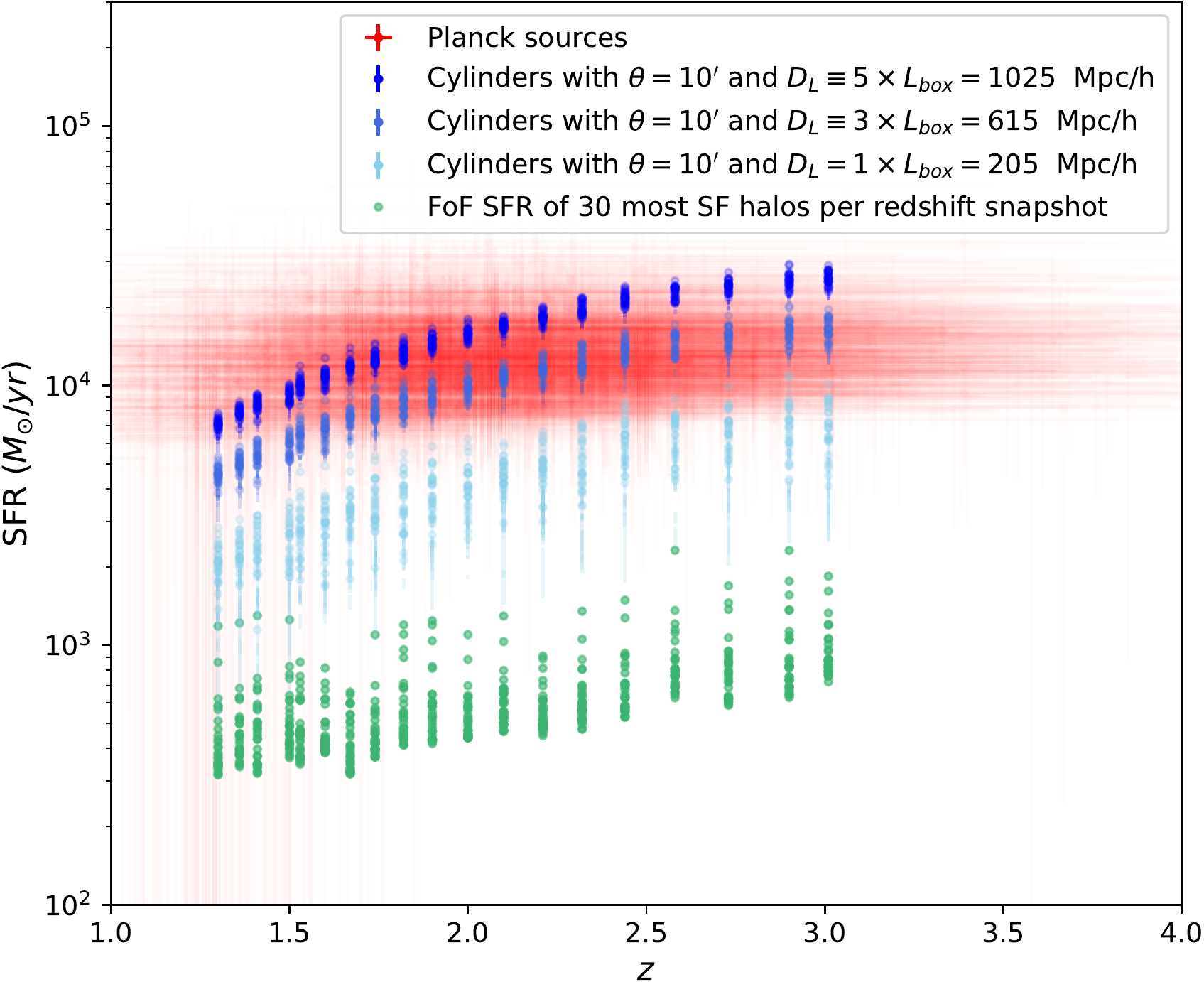}
     \caption{Estimated SFRs and redshift of the Planck high-$z$ sources \citep{Planck2016_proto} as measured using \plk data (red points). For comparison, the SFRs of the 30 most SF simulated halos from 19 snapshots at different redshifts are also shown. In blue colors, we show the simulated halos SFRs computed by summing the SFR of all galaxies inside cylinders with diameter $\theta$\,=\,10 arcmin and lengths of $D_{L}=$ 205, 615 and 1025 Mpc/h (see Sect. \ref{SEC:METHOD}), and in green the FoF SFR values. \label{fig:PLANCK_SFR_z}}
\end{figure}

As discussed in Sect. \ref{SEC:DATA}, the \plk SFRs of the PHz sources are obtained by integrating the flux densities over a ${\sim}10$ arcmin region, and are expected to be highly over-estimated due to the contribution of SF galaxies along the LOS~\citep{Negrello2017}. Indeed, in spite of the cleaning procedure applied to remove the contribution to the sub-mm flux densities from low ($z{<}$1) and high ($z{>}$4) redshift sources, the range of possible redshifts of these \plk sources remains quite large, $\delta_z \sim 1.4$ (see Fig. \ref{fig:PLANCK_deltaz_z}).

To compare the \plk SFR with those obtained from the simulations, we consider the most SF halos at $1.3<z<3$ (in Sect. \ref{SEC:METHOD}) and compute their \plk-like SFR by considering cylinders of diameter $\theta$\,=\,10 arcmin, and of lengths $D_L$\,=\,$205$, $615$, and $1025$ cMpc/h.
The total SFRs obtained from the simulations and from \plk \citep{Planck2016_proto} are shown in Fig. \ref{fig:PLANCK_SFR_z} with, respectively, red, and blue symbols. 
The FoF SFRs of the simulated most SF halos, derived by adding the SFRs of all the bounded galaxies, are also shown as green circles.
Notice that the \plk SFRs have been divided by 1.74 to be compared with simulations, to correct from the Salpeter IMF \citep{Salpeter1955} assumed by \plk to the adopted Chabrier IMF \citep{Chabrier2003} in simulation.

The SFRs obtained by considering all simulated galaxies inside cylinders with length $D_L$ from $615$ up to $1025$ cMpc/h, and aperture $\theta$\,=\,10 arcmin centered on the most SF simulated objects well match the SFRs derived from the \plk photometric measurements. 
On the contrary, these \plk-like integrated SFRs are about $25$ larger than the SFRs of most SF halos derived from the SFRs of the bounded galaxies. 
Thus we argue that the SFR estimations from the \plk high frequency maps are the result of the sum of both, at least one very luminous source, and a large contamination of foreground and background sources along the LOS projected within a region similar to the \plk beam.

Interesting the length of $D_L$\,=\,$1025$ cMpc/h corresponds approximately to the photometric \plk redshift range $\delta_z$, and is also well matching with the effective SFR of Planck sources. Indeed, this amount of LOS contamination corresponds the associated photometric redshift range ($\delta_z \sim 1.4$; see Fig. \ref{fig:PLANCK_deltaz_z}) as determined by fitting a single modified blackbody to the \plk high frequencies flux densities. This similarity suggests that the uncertainty derived from the SED fitting is related to the contribution of IR sources along the LOS.
We can also notice that the effective SFR of Planck sources is matching with LOS contamination from 615 to 1025 cMpc/h, whereas their redshift uncertainties is better represented by LOS distances from 1025 to 1435 cMpc/h. 
This difference can be interpreted as a hint of an over-estimation of the Planck redshift uncertainties.

This finding which supposes that the effective flux (and derived SFR) of the \plk sources are over-estimated due to line of sight effects from aligned SF galaxies integrated in a large redshift range is consistent with both theoretical and observational studies of \plk sources.
Theoretically, \cite{Negrello2017} have shown that protocluster candidates detected in the \plk maps can be interpreted as Poisson fluctuations of the number of high-$z$ dusty protoclusters within the same Planck beam, rather then being individual clumps of physically bound galaxies. They have concluded that most of the flux density within the \plk beam can be explained by one or two very luminous sources, and by a larger number of faint galaxies along the LOS.
Moreover, follow-up observations have also revealed that these fields often contain two SF structures aligned along the same LOS, as illustrated in Fig. \ref{fig:DATA_FOLLOWUP}. 
As discussed in \cite{Polletta+21}, the total SFR from the SF galaxies, identified spectroscopically or through narrow-band imaging, in a small redshift range ($z=[2.15-2.2]$) is much smaller than the SFR derived from \plk sub-mm measurements, implying the need of integrating the signal along a larger redshift range to reproduce the \plk-derived SFR of the PHz sources. Indeed, they reproduce the \plk flux densities measured in the \plk beam of a PHz source by summing the Herschel flux densities of all the Herschel sources in the same region.


\subsection{Comparison with PHz follow-up observations}
\begin{figure*}
    \centering
    \includegraphics[width=0.71\textwidth]{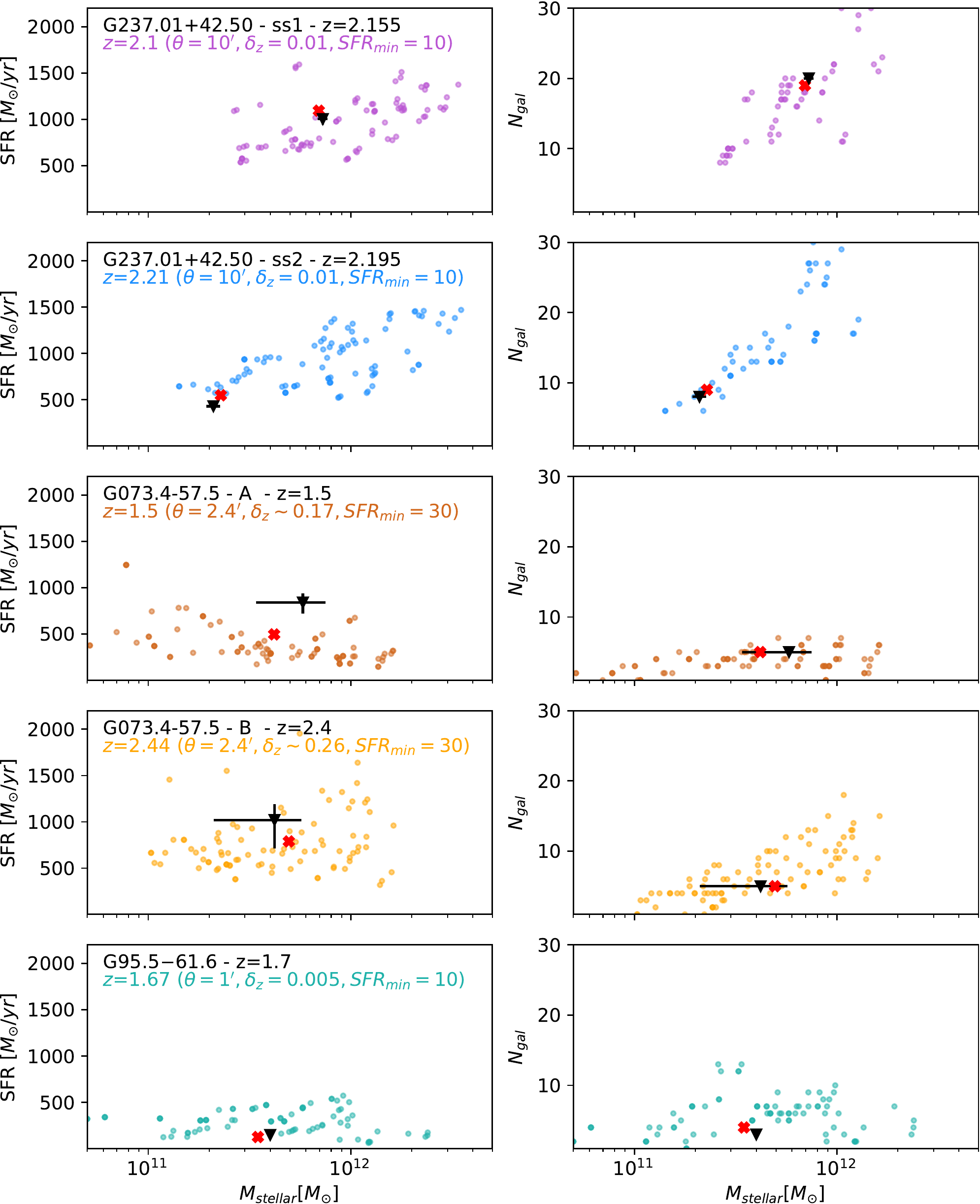}
     \caption{Total SFR, and stellar masses (\textit{left panels}), and number of galaxies (\textit{right panels}) derived from the 30 most SF simulated objects at specific redshifts, $z_{snap}$ compared with the measured values from five structures (one per row) found in three PHz sources. The selected snapshot in each row is the closest to one of the observed PHz structures. The reported quantities have been obtained by averaging the total values measured in three cylinders (oriented along x, y, and z-axis) with parameters $\theta$, and $\delta_z$, and considering only galaxies with $SFR{>}SFR_{min}$. 
     Each row corresponds to one of the five observed structures (the name and the redshift are noted in black), and relative quantities (see Table~\ref{tab:DATA}) are shown with a black upside-down triangle. The snapshot redshift, the SFR threshold, and the cylinder parameters are noted in each panel in various colors, and shown as color circles. A red cross corresponds to the simulated halo with estimated quantities that best reproduce those of an observed structure.}
     \label{fig:follow_up_1}
\end{figure*}
\begin{table*}[]
    \centering
    \caption{Properties of simulated halos that best reproduce the observed values}
    \begin{tabular}{l c c c c c c c c c}
    \hline\hline
        Observational case & $z_{snap}$ & \multicolumn{3}{c}{cylinder parameters} & $N_{gal}$ & SFR             & $M_{*}$                   & $SFR^{FOF}$ \\
                           &            &    $\theta$  & $\delta_z$ & $SFR_{min}$ &           & [$M_{\odot}/yr$]& [$10^{11} \ M_{\odot}/h$] &  [$M_{\odot}/yr$] \\
        \hline
        G237.01+42.50 \textit{ss1} & 2.1  & 10$^{\prime}$  & 0.01  & 10   & 19 & 1092 &  6.9   & 654 \\
        G237.01+42.50 \textit{ss1} & 2.21 & 10$^{\prime}$  & 0.01  & 10   &  9 &  548 &  2.3   & 490 \\
        G073.4-57.5 \textit{A}     & 1.5  & 2.4$^{\prime}$ & 0.17  & 30   &  5 &  495 &  4.1   & 368 \\
        G073.4-57.5 \textit{B}     & 2.44 & 2.4$^{\prime}$ & 0.26  & 30   &  5 &  788 &  4.9   & 712 \\
        G95.5-61.6        & 1.67 & 1.0$^{\prime}$ & 0.005 & 10   &  4 &  125 &  3.4   & 413 \\
        \hline
    \end{tabular}
    \tablefoot{The properties of most SF halos that are the closest to those of the observed PHz structures, in term of their stellar mass, SFR, and number of SF galaxies computed from cylinder parametric model  (shown as red stars in Fig. \ref{fig:follow_up_1}).
    \label{tab:SIMU}}
\end{table*}

Three PHz sources have been followed-up with dedicated observations that have yielded the discovery of significant overdensities of galaxies at similar redshifts confirming their association with high-$z$ protoclusters, as described in Sect.~\ref{SEC:DATA} and summarized in Tab.~\ref{tab:DATA}.
These PHz fields, G237.01+42.50, G073.4-57.5, and G95.5-61.6, have been individually observed with distinct observational strategies (spatial and wavelength coverage, redshift measurements techniques, and depth), making direct comparisons with simulations difficult. 
More specifically, the overdensity in the PHz source G237.01+42.50 has been found by combining spectroscopic observations at optical and near-IR wavelengths over a $10'\times10'$ region. These observations were biased in favor of SF galaxies with sufficiently bright emission lines at rest-frame ultraviolet or optical wavelengths to enable a spectroscopic redshift measurement~\citep{Polletta+21}. Once an overdensity at a specific redshift was found, narrow-band imaging observations in a smaller region of the field were carried out to identify SF galaxies with a strong H$\alpha$ line in emission at the same redshift~\citep{Koyama2021}. The PHz source G073.4-57.5 has been instead observed with ALMA pointed observations targeting eight Herschel sources in a $5'\times5'$ region~\citep{Kneissl+19}. Finally, in G95.5-61.6 optical-NIR spectroscopic observations were carried out targeting color-selected galaxies associated with four Herschel sources, all located in a $1'\times1'$ region~\citep{Flores-Cacho+16}.

In this second comparison, our goal is to test whether the galaxies in and around the most SF high-$z$ halos in IllustrisTNG can reproduce the galaxy properties observed in the overdensities found in these three PHz sources.
For each overdensity, we search for the best matched simulated halo by choosing the 30 most SF halos at the closest redshift, and for each simulated halo by selecting the galaxies using our parametric cylinder model. 
The parameters of the cylinder are chosen to be close to the observational configuration of each PHz observations, and reported here:

\begin{itemize}
    \item Two overdensities have been found in PHz G237.01+42.50, both in the same $10 \times 10$ arcmin$^2$ region, one containing 20 galaxies at $z=2.155\pm 0.005$ (\textit{ss1}) and anther with 8 galaxies at $z=2.195\pm 0.005$ (\textit{ss2})~\citep{Polletta+21}. We therefore consider cylinders with parameters $\theta=10$ arcmin and $\delta_z = 0.01$. 
    In addition, since the identified members have SFRs $> 10 M_{\odot}/yr$, only simulated galaxies with SFRs above such a value have been taken into account.
    \item The 8 ALMA pointings in PHz G073.4-57.5 covered, in total, an area of 2.4 arcmin$^2$ and yielded mm continuum detections of 18 galaxies. Photometric or CO-based spectroscopic redshifts were derived for these galaxies suggesting the existence of two overdensities, both with five members, one at $z \sim 1.5$ and another at $z \sim 2.4$ \citep{Kneissl+19}. Since the ALMA pointings and the members of each overdensities are distributed on a $5'\times5'$ region, but only 10\% of it was probed by the follow-up observations, to match these overdensities we   consider cylinders with $\theta = 2.4$ arcmin and a sufficiently large depth along the line-of-sight to mimic the significant photo-$z$ error  (i.e.; $\delta_z=0.17$ at  $z \sim 1.5$ and $\delta_z=0.26$ at  $z \sim 2.4$).
    We consider also a SFR threshold of $30 M_{\odot}/yr$, given the lowest SFR value measured in these overdensities (i.e.; 44$^{+24}_{-16}$ M$_{\odot}$ yr$^{-1}$). 
    \item The third PHz that was spectroscopically confirmed is G95.5-61.6. This source contains two structures, one with three members at $z\sim1.7$ and a second (blended) with six members at $z \sim 2$, both distributed over a $1'\times1'$ region~\citep{Flores-Cacho+16}. Since there are no accurate SFR estimates for the structure at $z\sim 2$, in the following, we will consider only the structure at $z\sim1.7$. Given the tiny aperture and the precision of the spectroscopic measurements, we choose a cylinder with angular aperture $\theta=1$ arcmin and depth equivalent to a redshift interval of $\delta_z=0.005$. Only galaxies with $SFR >10 M_{\odot} / yr$ are considered.
\end{itemize}

For each of the five confirmed structures, we consider the 30 most SF halos at the snapshot redshift closest to the structure redshift. We compute their total SFR and stellar mass considering only the galaxies with SFR larger than the threshold defined above inside the appropriate cylinder.
By considering three orientations (along x- y- and z- axis) for each cylinder, we artificially increase our simulated sample of 30 most SF halos to 90 SF halos in each case.
The distribution of total SFRs and stellar masses for the simulated halos are compared with the measured ones in the left panels of Fig.~\ref{fig:follow_up_1}. Each row refers to one of the five observed structures. 
We also show, on the right panels of Fig.~\ref{fig:follow_up_1}, the number of galaxies inside   cylinders centered on each of the 30 most SF halos as a function of total stellar mass.
Black symbols refer to the observational results of the five observed structures, and color symbols represent the values drawn from the simulations. 

As shown in Fig. \ref{fig:follow_up_1}, our cylinder parametrisation yields SFRs, stellar masses, and galaxy counts in the simulated objects that are compatible with the values measured in the five observed PHz structures.
This is particularly true for G237.01+42.50-\textit{ss1} at $z\sim2.1$, G073.4-57.5-\textit{B} at $z\sim2.4$, and G073.4-57.5 at $z=1.7$. To better illustrate the good match between the observations and the simulations, we highlight with a red star the simulated object that is in closer agreement with the corresponding observed structure. The values of these simulated halos are written in Table~\ref{tab:SIMU} for comparison with the observational ones reported in Table~\ref{tab:DATA}, showing the good agreement between the observed structures and the simulated SF halos.
Interestingly, the most similar simulated object in case of G237.01+42.50-\textit{ss2}, is located on the tail of the SFR-stellar mass distribution of the simulated SF halos, suggesting that this structure is one of the least massive and less SF objects over the most SF halos at its redshift. The comparison with a such large sample of simulated halos shows also that the G073.4-57.5-\textit{A} structure has a larger SFR than predicted by the simulations, even if the total stellar mass and the number galaxy count are well reproduced by the simulations. This discrepancy could be explained if some of the galaxy members had higher than predicted SFRs, but similar stellar masses. Since the identified members of this structure are all bright mm sources, they are biased in favor of galaxies with large SFRs. Such a bias was not included in the choice of the closest simulated object.

We argue that the good agreement between simulations and observations is due to the cylinder parametric model, that allows us to take into account both the projected spatial distribution of the galaxies on the sky, and their distribution along the LOS, as derived from the observations.  
Taking into account only the contribution from the bound galaxies to each simulated halo would result in larger discrepancies between simulated and observed values (see e.g. the FoF SFR values in Table~\ref{tab:SIMU}).


\subsection{Galaxy properties in PHz follow-up observations}

\begin{figure*}
    \centering
    \includegraphics[width=0.95\textwidth]{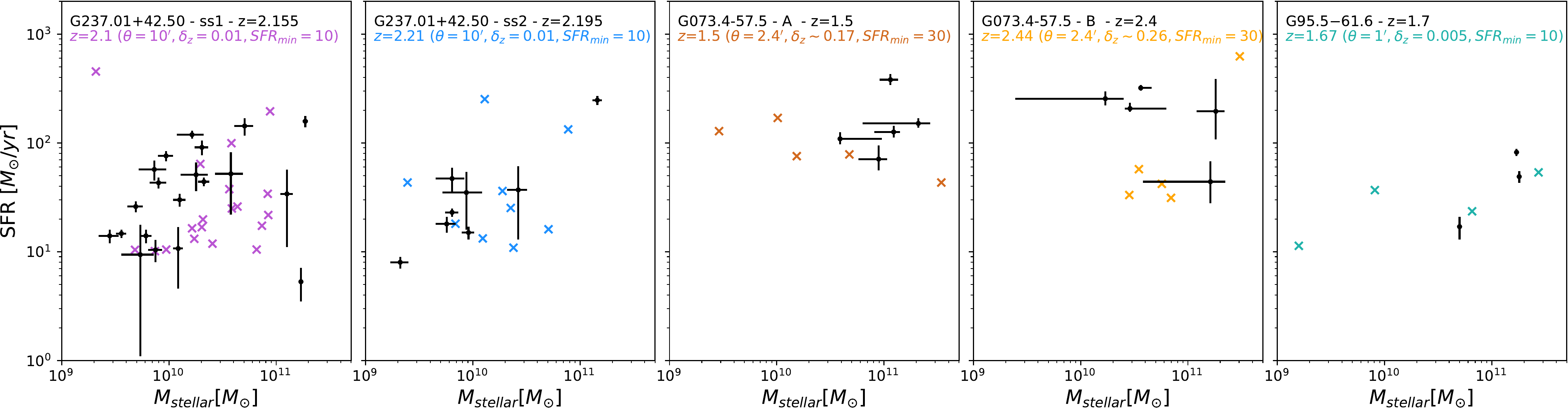}
     \caption{Star formation rates as a function of stellar masses of the galaxy members in the five confirmed structures (color crosses) and of those drawn from the closest simulated case (black points). The simulated data are obtained from all the galaxies inside a cylinder around a specific SF halo with total SFR and stellar mass that are the closest to those measured in the observations (see red cross in Fig. \ref{fig:follow_up_1}). 
     Each panel shows a different structure whose name and redshift are noted in black on the top. 
     \label{fig:follow_up_2}}
\end{figure*}
\begin{figure}
    \centering
    \includegraphics[width=0.45\textwidth]{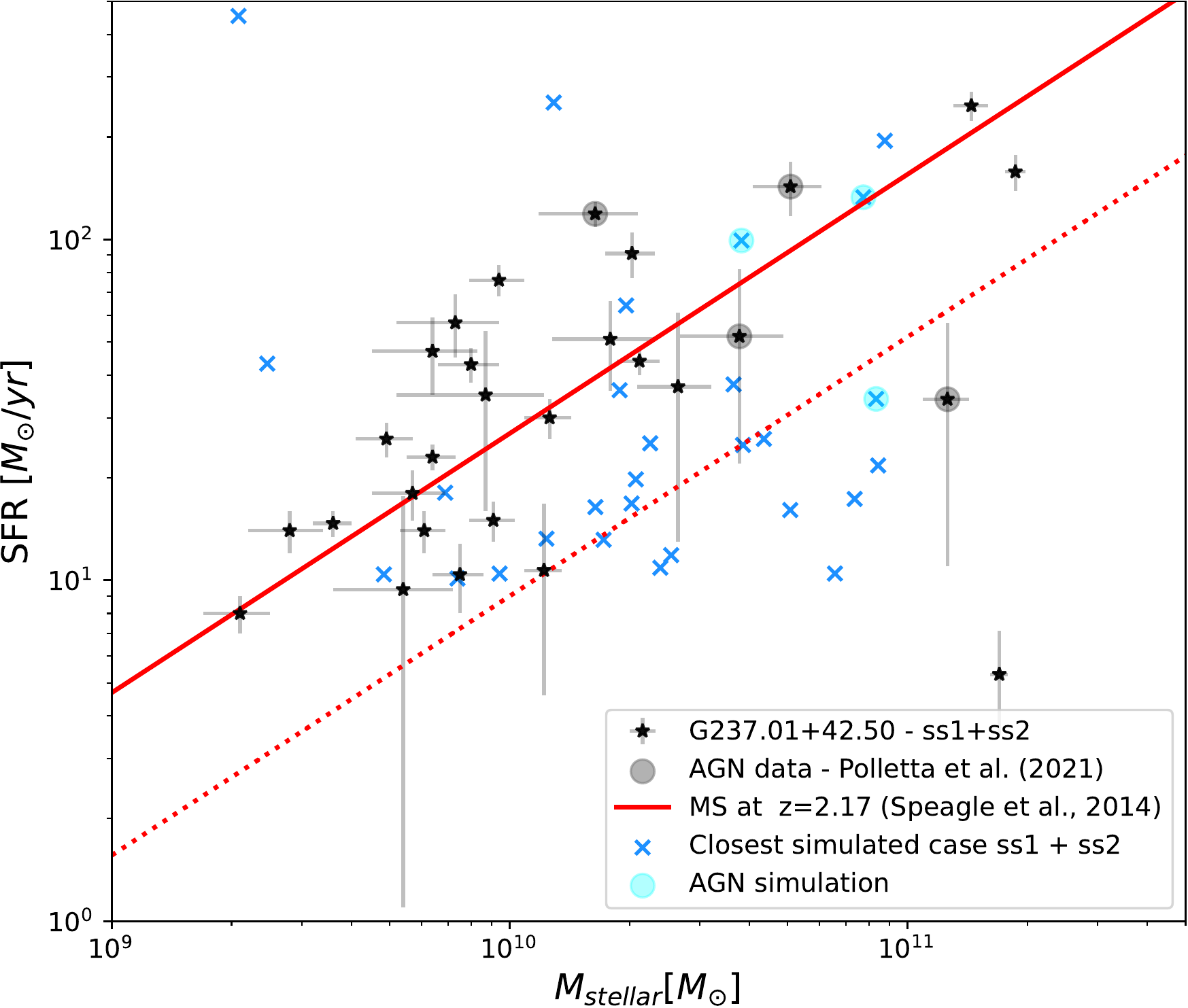}
     \caption{Star formation rates as a function of stellar masses of the galaxy members in the spectroscopically confirmed structure G237+42.50 (\textit{ss1}+\textit{ss2}) from \cite{Polletta+21} (black stars). A gray circle indicates the AGN members. The SFR and stellar masses of the galaxies from the closest simulated case (for \textit{ss1} and \textit{ss2}) are shown as blue crosses, and a cyan circle is over-plotted on those that are considered AGN. The main sequence (MS) at $z=2.17$ as formulated by \cite{Speagle2014} is shown as a solid red line, and the MS divided by a factor of $3$ as a red dotted line.
     \label{fig:follow_up_G0237}}
\end{figure}

The good agreement between the total SFRs and stellar masses of the simulated objects and the observed ones, prompts us to investigate whether this is also the case for the individual galaxy members.

To explore this point, we show in Fig.~\ref{fig:follow_up_2} the SFR and stellar mass of the galaxies in each observed structure (one panel per structure) as measured in the observations (black symbols), and in the simulated halo that best matches the integrated observed quantities (shown as red crosses in Fig.~\ref{fig:follow_up_1}; and in color crosses in Fig.~\ref{fig:follow_up_2}). 
The overall distribution of stellar mass and SFR of the galaxy members in the simulated SF halos and in the observed structures are consistent, in particular for the PHz G237.01+42.50 \textit{ss1} and \textit{ss2} structures for which a large number of galaxy members are known. Notice that galaxies inside the G073.4-57.5 \textit{A} and \textit{B} structures are apparently more massive (in the \textit{A} structure) and have higher SFRs (in the \textit{B} structure) than the simulated ones. Given that the membership is based on photometric redshifts (less accurate than spectroscopic redshifts), we believe that the comparison with the cylindrical model might not be well adapted, and a proper light-cone galaxy selection might be better to interpret these observations \citep[similar to][]{Araya2021}.
Also in G95.5-61.6, the simulated galaxies tend to be less massive and more numerous than the simulated ones, but this difference is not significant considering the small number of identified members.

We thus focus on the structure that has a large number of identified members, PHz G237.01+42.50. Since its two structures overlap in projection on the sky and are in relatively close proximity along the LOS, we consider them together for the following analysis. In Fig.~\ref{fig:follow_up_G0237}, we show the SFR as a
function of stellar mass of the 28 spectroscopic members of the two structures \textit{ss1} and \textit{ss2} \citep[as in Fig.~16 of][]{Polletta+21}, similarly to Fig.~\ref{fig:follow_up_2}.  We also show the star forming main sequence (MS) at $z$\,=\,2.17, as parametrized by~\citet{Speagle2014}.  Although these two structures can be considered as two distinct protoclusters, their close redshifts imply that they might constitute a proto-supercluster~\citep[see ][for a discussion on this possibility]{Polletta+21}.  As illustrated in
Fig.~\ref{fig:follow_up_G0237}, the observed galaxy members are consistent with the MS (assuming a \cite{Chabrier2003} IMF).  The galaxy members include both star-forming galaxies and Active Galactic Nuclei (AGN).  The latter have been identified through optical spectroscopy or X-ray data.  The estimated AGN fraction in the two combined strutures, PHz G237.01+42.50 \textit{ss1}, and \textit{ss2}, is $14\%\pm10\%$.

For comparison, we show in Fig.~\ref{fig:follow_up_G0237} the galaxies in the two simulated halos that best reproduce the integrated properties of the two structures in G237.01+42.50 (see red crosses in Fig.  \ref{fig:follow_up_1}) as blue crosses.  In the simulations, it is possible to identify the galaxies that contain an AGN as those that host a super massive and fast accreting black hole (SMBH; with masses $M_{BH}>10^8 M_{\odot}/h$ and instantaneous accretion rate $\dot{M}_{BH}>10^8 M_{\odot}/h/(0.978Gyr/h)$).
These threshold are such that the BHs would double their mass within roughly 1 Gyr. It represents the most luminous AGNs in massive galaxies, such as their bolometric luminosity is around $ L_{AGN} \gtrsim 10^{44} \ ergs/s$, as estimated from BH model of \cite{Churazov2005}  \citep[we refer to][for details on AGN evolution and their bolometric and X-ray luminosities in TNG300]{Florez2021,Habouzit2022}.
We find an AGN fraction in the simulated objects that is $11\%\pm6\%$ (with two AGN in the simulated object matching \textit{ss1} and one in that matching \textit{ss2}).  The AGN fraction in the simulated \textit{ss1} and \textit{ss2} objects is thus consistent with the observed value.

Regarding the distribution of galaxies in the SFR-$M_{*}$ diagram, whereas the observed galaxies are well distributed around the MS, the simulated ones have almost systematically lower SFRs.
This result is consistent with the findings of \cite{DIANOGA2020}, that find that the SFRs of high-$z$ SF galaxies are under-predicted in the DIANOGA hydrodynamical simulation.  They show that this lower normalisation of the MS in simulations is stable against varying several subgrid and AGN feedback models. This offset of the MS for high-$z$ SF galaxies in simulation, has been explored in different recent numerical studies, and the reason is still debated.  A possible explanation might be linked to the underestimated gas fractions in high-$z$ galaxies~\citep{DIANOGA2020}, and to the spatial resolution limit in simulations~\citep{Lim2021}. 


\subsection{The fate of protocluster candidates}\label{SEC:FATE}

\begin{figure}
    \centering
    \includegraphics[width=0.45\textwidth]{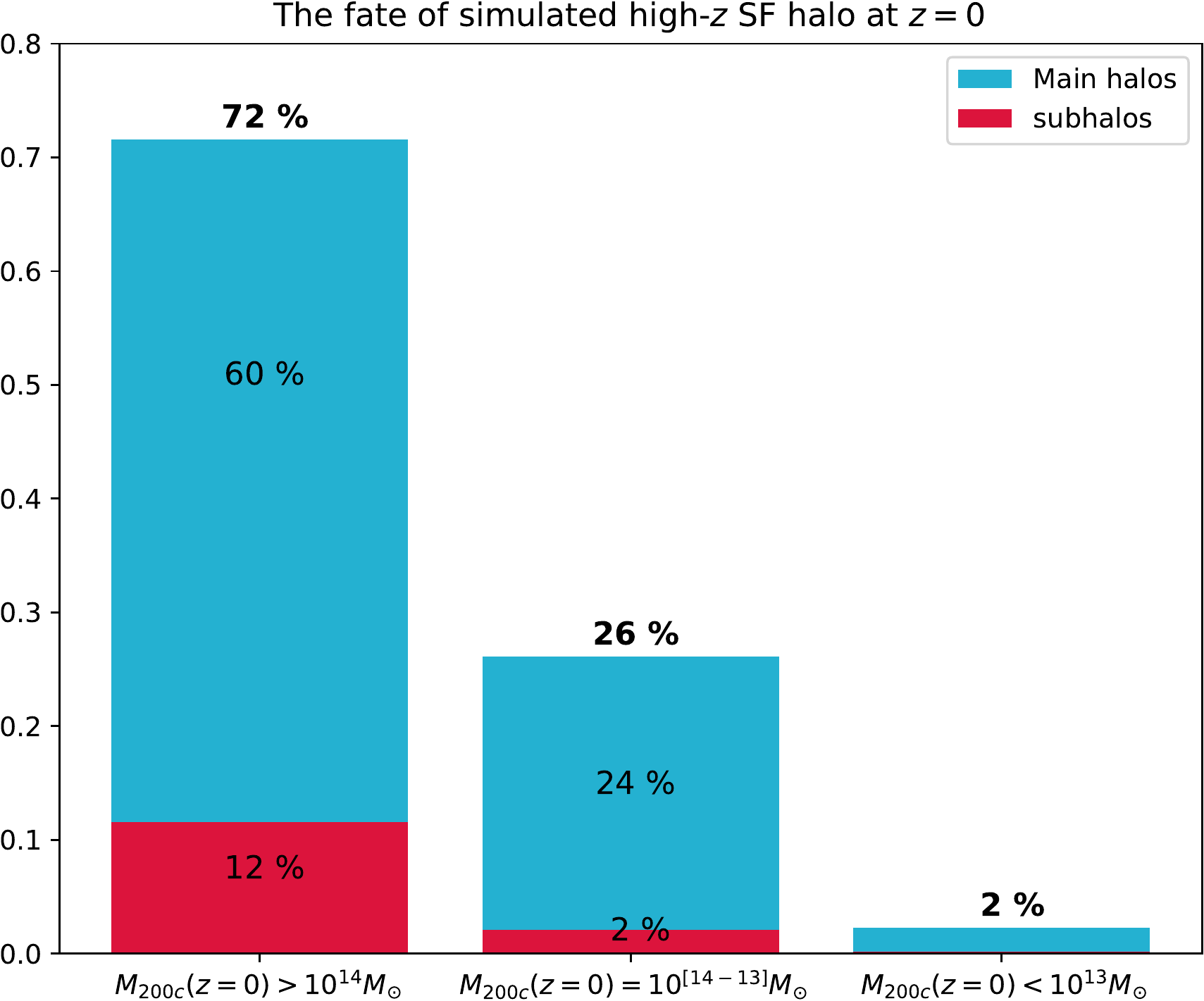}
     \caption{Fractions of simulated SF halos at high-$z$ that will evolve into clusters with halo mass $M_{200}>10^{14} M_{\odot}$, into groups with halo mass $M_{200}=[1-10]\times 10^{13} M_{\odot}$, and into single halos with $M_{200}< 1 \times 10^{13} M_{\odot}$ by $z$\,=\,0.
     We also show the fraction of simulated halos that will become the main subhalo of a halo at $z=0$ (turquoise), or a substructure (subhalo) inside a given halo (red).\label{fig:FATE}}
\end{figure}
\begin{figure}
    \centering
    \includegraphics[width=0.48\textwidth]{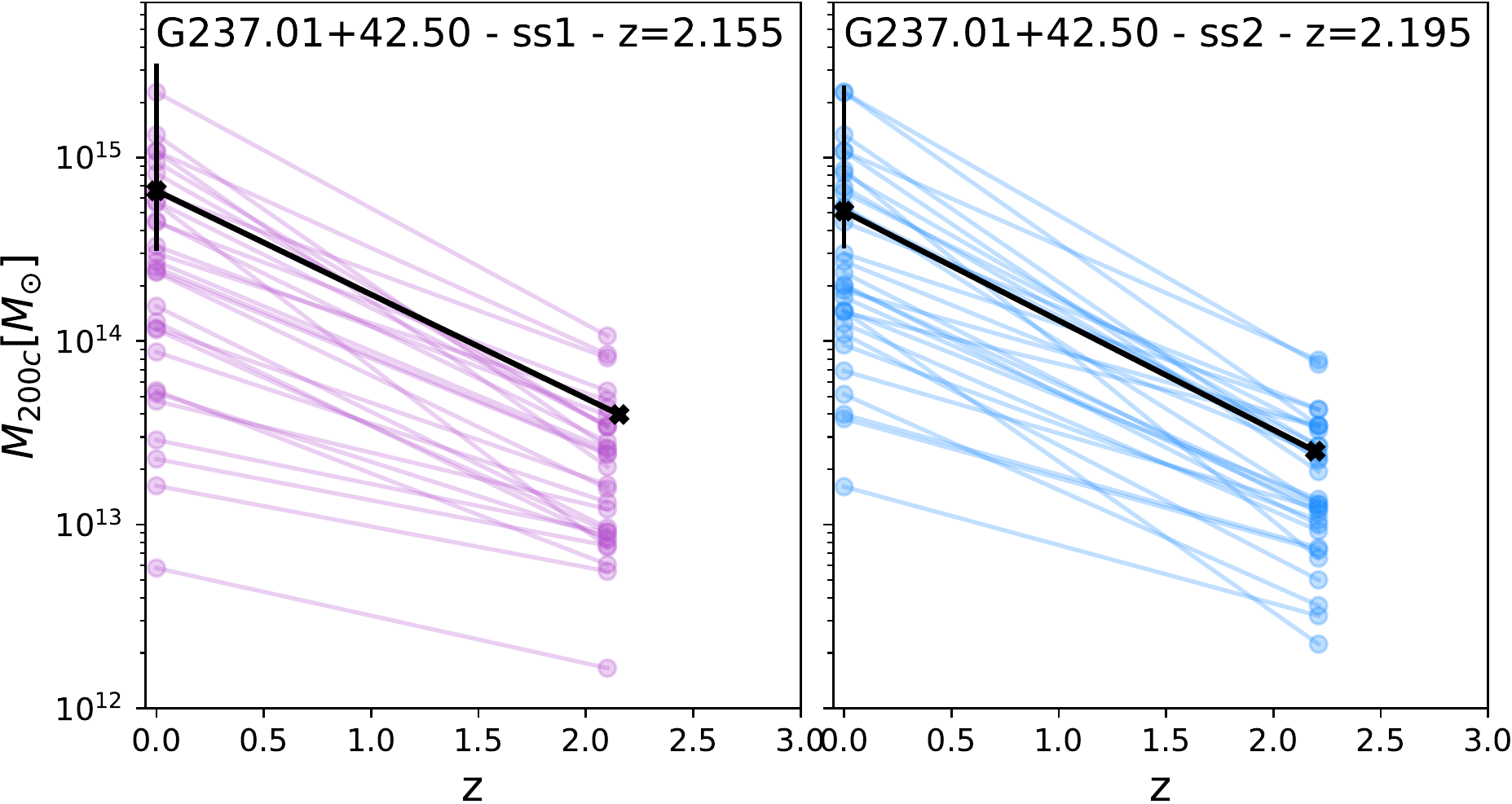}
     \caption{Halo mass evolution of the 30 most SF simulated halos at $z=2.1$ (purple circles connected by solid lines in the \textit{left panel}) and at $z=2.21$ (blue circles connected by solid lines in the \textit{right panel}), showing their mass at the observed redshift and the mass of their progenitor at $z=0$. The estimated halo mass of the two G237.01+42.50 \textit{ss1} and \textit{ss2} protoclusters and their expected mass at $z=0$ from ~\citet{Polletta+21} are shown in black. \label{fig:FATE_G237}}
\end{figure}

As described in Sect. \ref{SEC:METHOD}, we investigate the mass evolution of our 570 high-$z$ most-SF halo sample up to the present time at $z=0$. The results are presented in Fig. \ref{fig:FATE}. 
We define a SF halo as progenitor of a massive cluster if its mass at the present time is $M_{200}(z=0)>10^{14} \ M_{\odot}$ \citep{Chiang2017}.  We find that 72\% of our simulated protocluster candidate sample will actually become galaxy clusters by $z=0$.  The rest of the simulated sample contains predominately progenitors of galaxy groups with masses between $10^{13}<M_{200}<10^{14} M{\odot}$ (26\%), and a small minority (2\%) will evolve into low-mass ($M_{200}<10^{13} M{\odot}$) isolated objects at the present time.

Moreover, by considering separately main halos and subhalos inside a given halo at $z=0$, we find that 60\% of our SF simulated halo samples will become the main halo of clusters, whereas 12\% of them will become other satellite substructures inside clusters. For progenitors of group-size structures at the present time, we find that only a small fraction (2\%) are becoming substructures inside group-mass type objects. Finally, only 2\% of the most SF halo at high-$z$ will not merge into massive structures, but rather will stay as isolated low-mass halos at $z=0$.

Notice that a large number of our simulated halos will merge into the same structure by $z=0$. Indeed, the 570 SF halos at 1.3${<}z{<}$3 will yield 253 distinct halos at $z=0$ (or 279 distinct subhalos), given that they can merge or be the direct descendant from a snapshot to another at lower $z$. The evolutionary connection between the simulated halos drawn from different snapshots is due to our selection method.  Indeed, by performing a SFR-based selection at each snapshot from $z=1.3$ to $z=3$, we do not distinguish if the selected halos are direct descendant from one snapshot to the next one. Thus, by considering the progenitor at different redshifts of the same final structure we might bias our statistics on the fate of our simulated halo sample. We have tested this issue in Appendix \ref{APP:fate}, and find that the final percentage values are only slightly affected. The number of cluster progenitors decreases from $72\%$ to $63\%$, and that of group progenitors increases from $26\%$ to $33\%$.  Thus, the result that the vast majority of the most SF halos at high-$z$ will evolve into massive clusters by $z$\,=\,0 remains valid.

\begin{figure*}
    \centering
    \includegraphics[width=0.9\textwidth]{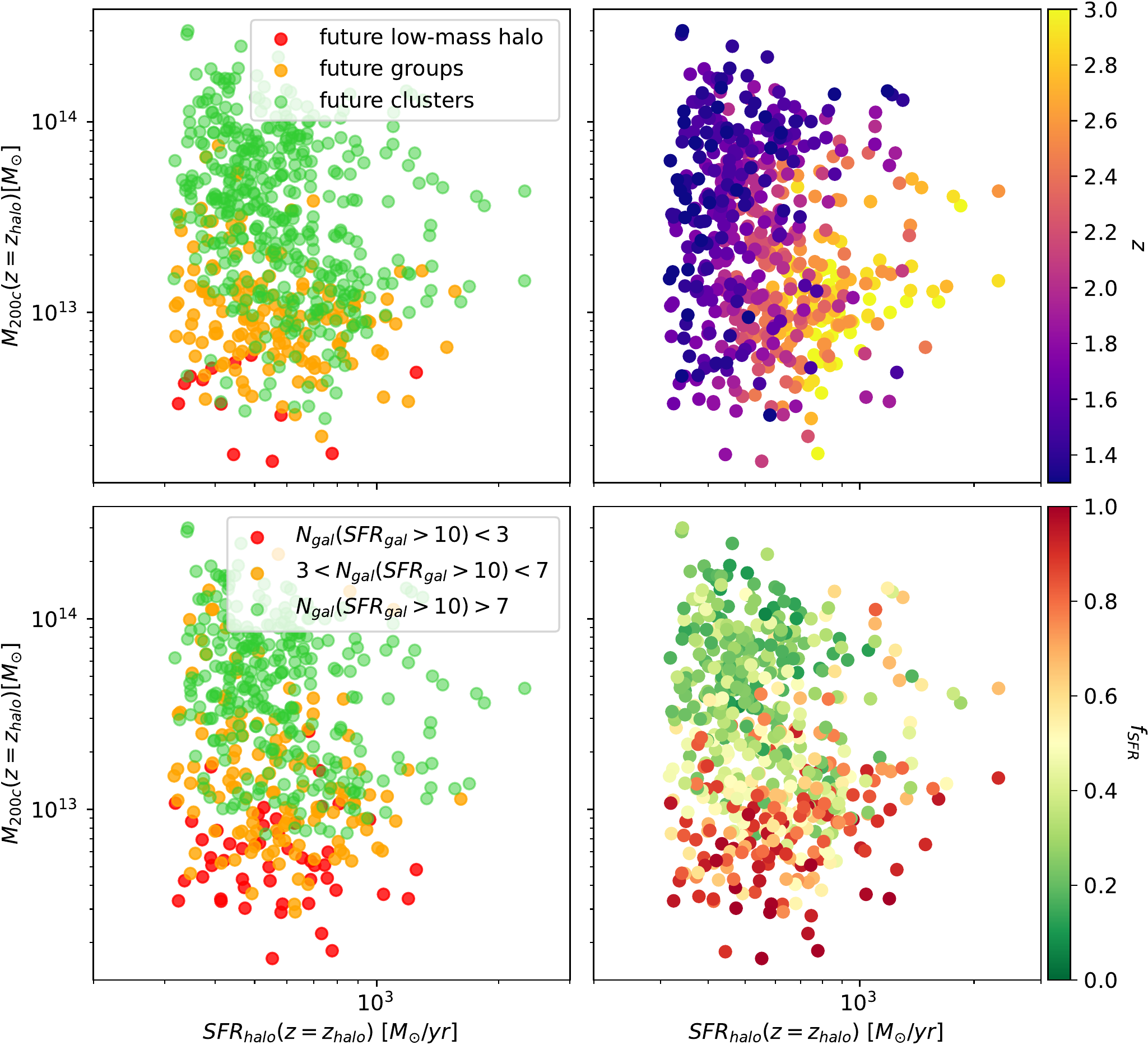}
     \caption{The distribution of mass $M_{200c}$ and SFR of our sample of 570 simulated high-$z$ SF halos, colored by their fate at $z=0$ (left top panel), their redshift (right top panel), their number of SF galaxy member with galaxies are $SFR>10 \ M_{\odot}/yr$ (left bottom panel), their SFR fraction (as defined in Eq.~\ref{eq:frac_sfr}, right bottom panel).  \label{fig:FATE_SFR}}
\end{figure*}

To assess the possible range of estimated final masses of a high-$z$ simulated halo, we show the predicted $z$=0 masses of the 30 most SF halos of our simulated sample at $z=2.1$ (right panel) and at $2.21$ (left panel) in Fig.~\ref{fig:FATE_G237}. We consider these two redshifts because they match those of the two structures in G237.01+42.50, and we find a good agreement between the observed galaxies in the two  structures and the simulated ones selected via our cylinder model (see Fig.~\ref{fig:follow_up_G0237}). One might ask what will be the fate of the most SF simulated halo at a redshift similar to that of the G237.01+42.50 \textit{ss1} and \textit{ss2} protoclusters. In Fig.~\ref{fig:FATE_G237}, we show the halo mass of the most SF halos at $z=2.1$ and $z=2.21$ and their final masses at $z=0$. We overplot with black points the halo masses of G237.01+42.50 \textit{ss1} and \textit{ss2}, and their expected mass at the present time as deduced from the analytical method of \cite{Steidel1998} using the galaxy overdensity, as derived by \cite{Polletta+21}. The theoretically expected fate of \textit{ss1} and \textit{ss2} structures is consistent with the fate of the most SF halos at their own redshift in hydrodynamical simulation. 
The large scatter of halo masses from high-$z$ to $z=0$ reflects the wide spread in accretion history of structure building-up \citep[as discussed in][]{Rennehan2020}.
We estimate a probability of 73\% for \textit{ss1}, and 80\% for \textit{ss2} to become massive clusters with a mass larger than $1\times 10^{14} M_{\odot}$ at $z=0$.

\subsection{Which protocluster property best predicts its fate?}

Making use of the simulations, we explore whether there is an observable in protoclusters at high-$z$ that can give hints of their fate at present times. For this analysis, we consider the  intrinsic halo properties at the observed $z_{halo}$, as defined by the FoF catalog, such as the halo mass $M_{200}$, the redshift, the FoF SFR, and the number of galaxy members~\citep[defined as the hosted subhalos within each FoF halo;][]{ILLUSTRIS_TNG}.  For each halo, we also define a SFR fraction as: 
\begin{equation}
    f_{SFR} = \frac{\max (SFR_i) }{SFR_{halo}} \,,
    \label{eq:frac_sfr}
\end{equation}
where $N_{gal}$ is the number of galaxies gravitationally bound to a given halo based on the FoF algorithm. For each $i$ galaxy associated to a given FoF halo, we note its star formation rate $SFR_i$. The SFR fraction $f_{SFR}$ thus defines the contribution of the most SF galaxy to the total SFR of its parent halo.

We also quantify the number of galaxies per halo using a SFR threshold to remove low-SF galaxies, that would be challenging to spectroscopically identify as protocluster members in moderately deep observations.  We choose a SFR threshold $SFR_{gal}=10 M_{\odot} yr^{-1}$ -- as done to reproduce the observations discussed in this work.

 The results are presented in Fig.~\ref{fig:FATE_SFR}, where we show the halo mass at their redshift, $z_{halo}$ as a function of their total SFR for the full sample of 570 simulated halos. 
 In each panel, we color code the symbols according to a specific property, the final halo mass at $z$\,=\,0 (top left panel), the redshift (top right panel), the number of SF galaxies with $SFR>10 \ M_{\odot} yr^{-1}$ (left bottom panel), and the SFR fraction $f_{SFR}$ (right bottom panel).

As we can see on the left top panel, the fate at $z=0$ of our SF halo sample is strongly governed by the halo mass at $z=z_{halo}$. 
As expected, a massive halo at high-$z$ is supposed to grow by accretion and merger to become a more massive structure at $z=0$, following hierarchical formation scenario.
Interestingly, the total SFR of a high-$z$ halo is not a good parameter to establish whether it will evolve into a massive cluster at $z=0$. Indeed, the final mass of a high-$z$ halo depends mainly on its mass at their redshift, and not on its total SFR.

One might question whether this is the result of our halo selection where we consider only the most SF halos from $z=1.3$ to $z=3$. In the right top panel, we probe the dependence of the halo properties on their redshift. 
We can see that the redshift of the simulated SF halos does not have an impact on their fate. On the other hand, there is relation between the halo redshift and the halo SFR, with halos at higher redshifts being more SF than those at lower $z$. As illustrated in Fig. \ref{fig:PLANCK_SFR_z}, and also shown over a wide (from $z=0$ up to $z=7$) redshift range in \cite{Lim2021}, the SFRs of cluster progenitors are supposed to peak at around $z\sim3-4$, and decrease from $z\sim 3$ to $z=0$.

We now examine whether the fate of the high-$z$ SF halos can be predicted based on their galaxy properties at the observed redshift (bottom panels of Fig \ref{fig:FATE_SFR}). For this analysis, we consider the number of galaxies above a minimum SFR of 10$M_{\odot} yr ^{-1}$ within each high-$z$ halo ($N_{gal}$; bottom left panel), and the SFR fraction ($f_{SFR}$; bottom right panel). The smaller the number of galaxy members or the higher the SFR fraction (meaning that the total SFR is dominated by a single galaxy), the more a high-$z$ SF halo leans towards an isolated low-mass structure at $z=0$. Conversely, high-$z$ halos with a more evenly distribution of SFRs over a relatively large number of SF galaxies (rather than having a single dominant galaxy with a very high SFR) are more likely to evolve into massive clusters at the present time. 

To investigate the relation between the global halo properties and those of the galaxy members, we analyze the halo mass - SFR distribution of the high-$z$ SF halos in three bins of the number of SF galaxy members ($N_{gal}$ below 3, between 3 and 10, and larger than 10; see different color symbols in the bottom left panel of Fig.~\ref{fig:FATE_SFR}).  This binning in SF galaxy number well reproduces the distribution in final mass of the high-$z$ halos (as shown by comparing the top and bottom left panels). The number of SF galaxies in a high-$z$ halo is, indeed, a strong indicator of its fate at $z=0$. Interestingly, this relation is independent of the total SFR of the halo, although we have to keep in mind that our sample might not probe a sufficiently broad range of total SFRs to show significant trends with it as it contains only the most SF halos at each redshift snapshot.
In summary, the more populated a SF halo is, independently of its redshift (from $z=1.3$ to $3$), the higher is the chance to become a massive cluster by $z=0$. More quantitatively, having more than seven SF galaxies gravitationally bound appears to be a strong hint to actually be a massive cluster-progenitor. Indeed, the probability for a high-$z$ SF halo with more than seven SF galaxies in our sample to be a massive cluster progenitor is about 92\%.

Such a result provides a powerful diagnostic to interpret SF protocluster candidates at high redshift ($1\lesssim z<3$). This is consistent with some theoretical models \citep[see e.g.][]{Steidel1998}, and argues in favor of overdensities of SF galaxies to trace the  most-massive dark matter structures at high-$z$ \citep{Cowley2016}. 
Furthermore, this is in line with \plk \citep{Planck2014_18} exhibiting a correlation between dark matter halos and the Cosmic Infrared Background, CIB \citep{Puget1996,Hauser2001,Dole2006}, in stacks, highlighting the relationship between dark matter halos and star-formation.
In general, the detection of a protocluster candidate is more reliable if it is supported by a high value of galaxy overdensity (galaxy density in a source in contrast to the galaxy density field). Our analysis suggests that indeed, rather than an extreme value of SFR from an individual high-$z$ galaxy, a large overdensity of SF galaxies is a more powerful indicator to find massive cluster progenitors at $z=0$.

\section{Discussion \label{SEC:DISC}}

With a new approach using state-of-the-art hydrodynamics simulations, we investigate the type of high-$z$ structures that are selected as bright and red sub-mm sources in the \plk maps, the so called PHz~\citep{Planck2016_proto,Planck2016}.  
We find that the high observed \plk sub-mm flux densities, and thus star formation rates, are reproduced by the simulations if multiple structures at 1.3${<}z{<}$3 along the LOS are taken into account.
However, most of the PHz sources seem to contain a high-$z$ star-forming structure that will evolve into a massive cluster by $z$\,=\,0, confirming their protocluster nature.
These results are consistent with previous findings, both from theoretical works~\citep{Negrello2017}, and from observations~\citep{Flores-Cacho+16,Kneissl+19,Koyama2021,Polletta+21,Polletta+22}.

Notably, we establish a new diagnostic to assess whether a high-$z$ ($z>1.3$) structure is a star-forming protocluster that will become a massive cluster by $z=0$. This can be gauged by combining the following observables: a) the number of star-forming galaxies in the structure (i.e., $N>7$) and b) the distribution of star-formation among all galaxy members (i.e., better a more even distribution than having a single highly star forming member; see Figs.~\ref{fig:FATE} and \ref{fig:FATE_SFR}). This diagnostic is easy to apply and extremely useful to select the most promising structures for further studies and for additional follow-up observations.

Comparisons between observed protoclusters and numerical predictions have been already attempted in the past, but remain extremely difficult given both observational and simulation aspects. 
Firstly, the observational selection and measurements are hardly reproducible in simulations, as these would require precise mock observations from light-cone construction (to accurately account for LOS  contributions), knowing the galaxy spatial and redshift distributions, creating mock images with all observational effects (e.g.; noise, depth, PSF, angular and sampling effects, among others). Secondly, the capacity of hydrodynamical simulations to accurately reproduce high SF galaxies at high-$z$ is still controversial. This could be due to the limited resolution of the simulations~\citep{Lim2021}, to the biases introduced by the SED fitting models~\citep{Nelson2021_SFR}, to the adopted dust model in the simulations, and to far-field blending effects~\citep{Lovell2021}.

Our cylindrical model is the first attempt to explain the \plk high-$z$ SF protocluster candidate sample by using SF galaxy distributions from hydrodynamical simulations (and not just mass-selected samples in simulations). Currently, there is no common established technique to estimate the SFR of galaxy protoclusters in simulations (either considering integrated aperture depending on the halo radius $R_{500}$ or a fixed aperture in comoving or physical scales). We thus developed a simple parametric model consisting in integrating galaxies inside a cylinder centered on the area of interest, i.e. the most SF halos. We argue that such a method provides a good agreement with \plk follow-up observations, and fairly reproduces the PHz protocluster measurements.
However, it is important to emphasize that a proper reproduction of the \plk selection would require the use of light-cone simulations, even if the number of protoclusters found in a specific redshift range is limited by the light-cone volume. Therefore, light-cones might provide a better solution to specifically assess the completeness and purity of cluster and protocluster detections by creating mock data images, and might thus be constructed over N-body simulations \citep{Blaizot2005,Ascaso2016,Krefting2020,Araya2021}. 
Exploring the complete projection of SF source distribution from $z=1.3$ and $3$ in hydrodynamical simulation and inside \plk beam size will be the next necessary step to accurately evaluate the number of sources and their individual contributions along the large LOS integration (as $\delta_z \sim 1.4$).

\section{Conclusions}
The main goal of this work was to investigate the nature of the \plk selected high-$z$ sources and the origin of their bright sub-mm flux densities, and high star formation rates. To this end, we have examined the spatial distribution and the properties of the SF galaxies associated with these bright \plk sources using state-of-the-art hydrodynamical simulations.
The PHz are detected in the \plk high frequency maps (over the cleanest 26\% of the sky), after removing Galactic dust emission (and sub-mm emission from sources at $z<1$), and CMB contamination (and emission from $z>4$ sources). The more than 2100 PHz sources with estimated redshifts around $z\sim2$, represent an ideal sample for investigating the most active sites of star formation  during the epoch of peak activity (i.e.; from $z\sim1$ to $3$), the so-called Cosmic Noon~\citep{Madau2014,Chiang2017}. 

In this work, we select a sample of $570$ high-$z$ SF objects representing the thirty most SF halos at 19 different redshifts from $z=1.3$ to $z=3$ in the TNG300 simulation of the IllustrisTNG project \citep{ILLUSTRIS_TNG}. This SFR-based selection technique provides a better representation of the observed protoclusters than other selections based for example on mass, as also shown by \cite{Lim2021}.
The properties of these simulated objects are computed by considering the galaxy distribution inside parametric cylinders centered on each SF halo, in order to reproduce the aperture window size of the observations, and possible LOS contaminations (integral in redshift range), an effect discussed by \cite{Negrello2017}. 
This cylindrical toy model has been designed to select the galaxies associated with each high-$z$ SF object in the simulations, to compute their total SFR, and stellar mass, and to characterize their galaxy member properties.
We compare the properties of the simulated high-$z$ SF objects with the \plk measurements derived for the whole PHz sample and with more detailed observations carried out for three PHz sources for which significant galaxy overdensities have been found through spectroscopic observations~\citep{Flores-Cacho+16,Kneissl+19,Polletta+21}. The results of this comparison are summarized below:

\begin{itemize}
    \item[1)] The total SFR of the PHz sources, measured from the \plk flux densities, can be reproduced in the simulations by taking into account the contribution of the most SF halos at a specific redshifts, and of the galaxies along the LOS distributed over a redshift interval consistent with a distance of $1025$ comoving Mpc/h. 
    This result implies a large contamination in the \plk sub-mm measurements from background and foreground galaxies along the LOS. 
    This LOS contamination is in agreement with \cite{Negrello2017} that find, from a semi-analytical analysis, that the high SF sources detected in the \plk maps can be interpreted as the sum of at least one high SF halo, and of a strong contamination from high-$z$ dusty galaxies along the LOS, both contributing within the \plk beam. 
    This finding is also consistent with results obtained from follow-up observations of \plk sources. The observations indeed reveal at least two distinct SF structures aligned along the LOS; and that the SFR derived from the \plk data can not be explained by the galaxies in one high-$z$ structure (suggesting the need of a significant contribution from galaxies along the LOS but projected within the \plk beam). 
    
    \item[2)] The number of galaxy members, and the total stellar mass and SFR of the spectroscopically confirmed structures found in three PHz sources through dedicated follow-up observations~\citep{Polletta+21,Kneissl+19,Flores-Cacho+16} are reproduced by the simulations. This good agreement is obtained by considering the most SF simulated halos at a redshift close to the observed one, and the SF galaxy population with a SFR above a certain limit and distributed within a cylinder of diameter and length consistent with the volume occupied by the confirmed members. Our cylinder parametric model can thus reproduce the PHz confirmed structures with high-$z$ SF simulated objects in terms of number of galaxy members, total stellar mass, and total SFR.
    
    \item[3)] In more details, comparing the values of SFR and stellar mass of the individual galaxy members in the observed structures and in the simulated SF objects gives a good agreement. In one case (i.e.; in G237.01+42.50), we were also able to test whether the fraction of AGN in the observed structure is consistent with that found in the simulation, finding a good agreement.
    The distribution of simulated galaxies with respect to the main sequence at their redshift is however shifted to lower values in SFR compared to that of the observed galaxies. This discrepancy is consistent with the results reported by~\cite{DIANOGA2020} where the normalisation of the main sequence at $z\sim 2.15$ is under-predicted by a factor of about $2-3$. This implies that, even if the total mass and SFR appear coherent with the data by integrating the galaxy distribution in a cylinder, the intrinsic under-prediction problem of galaxy SFRs in simulation at high-$z$ is not fully solved~\citep{Granato2015,Lim2021}. 

    \item[4)] Following these three last findings, we can conclude that our sample of the most SF simulated halos at high-$z$ is representative of the \plk sources when considering both the \plk measurements, and the spectroscopic observations. We can thus predict their fate at $z=0$, by probing the evolution of the simulated halo sample. We find that from 63\% to 72\% of our sample will actually become a massive galaxy clusters with $M_{200}>10^{14} M_{\odot}$ by $z=0$. It is important to point out that a significant, although minor, portion of them will evolve into a sub-structure inside a massive group or cluster ($\sim 15\%$). 
    
    \item[5)] One might ask which physical properties of these high-$z$ SF halos can give a hint on their fate at $z=0$. We find that, rather than the high value of SFR per galaxy, the number of SF galaxy members inside a halo (typically larger than $SFR>10 M_{\odot} yr^{-1}$) is an indicator of their evolution at $z=0$.
    In more details, high-$z$ halos which are populated by more than seven SF galaxies have a higher probability to be actual cluster progenitors. This finding argues in favor of overdensities of SF galaxies to trace the most massive dark matter structures at high-$z$ \citep{Cowley2016} and could be a new diagnostic to select high-$z$ protoclusters.
    
    \item[6)]By comparing simulations and observations based on a SFR selection, we confirm that the original \plk selection of PHz, despite LOS contamination, efficiently selects high-$z$ ($z \sim 2$) SF galaxy protoclusters, progenitors of $z=0$ massive clusters or substructures of clusters.  
\end{itemize}

As discussed in Sect. \ref{SEC:DISC}, this analysis was a first step towards explaining the \plk sources by considering the properties of the SF galaxy population in hydrodynamical simulations. To achieve a fairer comparison with the \plk  protocluster candidate selection and reproduce with fidelity the flux limit selection, it would be necessary to perform a light-cone analysis and create mock galaxy SEDs~\citep[similarly to][]{Araya2021}. 
Indeed a close agreement between our simulations and the observed SFR distribution from \plk is not expected because we do not exactly reproduce neither the selection criteria nor the SFR measurement procedure as adopted in~\cite{Planck2016_proto}. In conclusion, we confirm, from hydrodynamical simulation, the analytical finding from \cite{Negrello2017} about the interpretation of extremely high sub-mm flux from \plk sources as positive Poisson fluctuations of the number of high-$z$ dusty protoclusters within the same Planck beam.
The comparison with the spectroscopically confirmed structures argues in favor of a good agreement between observed SFRs and simulations using a parametric cylinder integral around the high-$z$ SF halos, but it also demonstrates that high-$z$ simulated galaxies suffer of SFR deficit. This result illustrates the long standing difficulty for numerical simulations to reproduce accurately the SFR of galaxies at the peak of the cosmic star formation rate density, the Cosmic Noon \citep{Granato2015,Dave2016,McCarthy2017,Donnari2019,Dave2019,DIANOGA2020,Lim2021,Yajima2022}.

\begin{acknowledgements}
The authors thank an anonymous referee for their useful comments and suggestions. This research has been supported by the funding for the ByoPiC project from the European Research Council (ERC) under the European Union’s Horizon 2020 research and innovation program grant agreement ERC-2015-AdG 695561 (ByoPiC, https://byopic.eu). 
The authors thank the very useful comments and discussions with all the members of the ByoPiC team.
We thank the IllustrisTNG collaboration for providing free access to the data used in this work. 
CG is supported by a KIAS Individual Grant (PG085001) at Korea Institute for Advanced Study.
\end{acknowledgements}

\bibliographystyle{aa}

\appendix

\section{Testing the main progenitor problem on the statistics of the fate of SF high-$z$ halos \label{APP:fate}}

\begin{figure}
    \centering
    \includegraphics[width=0.45\textwidth]{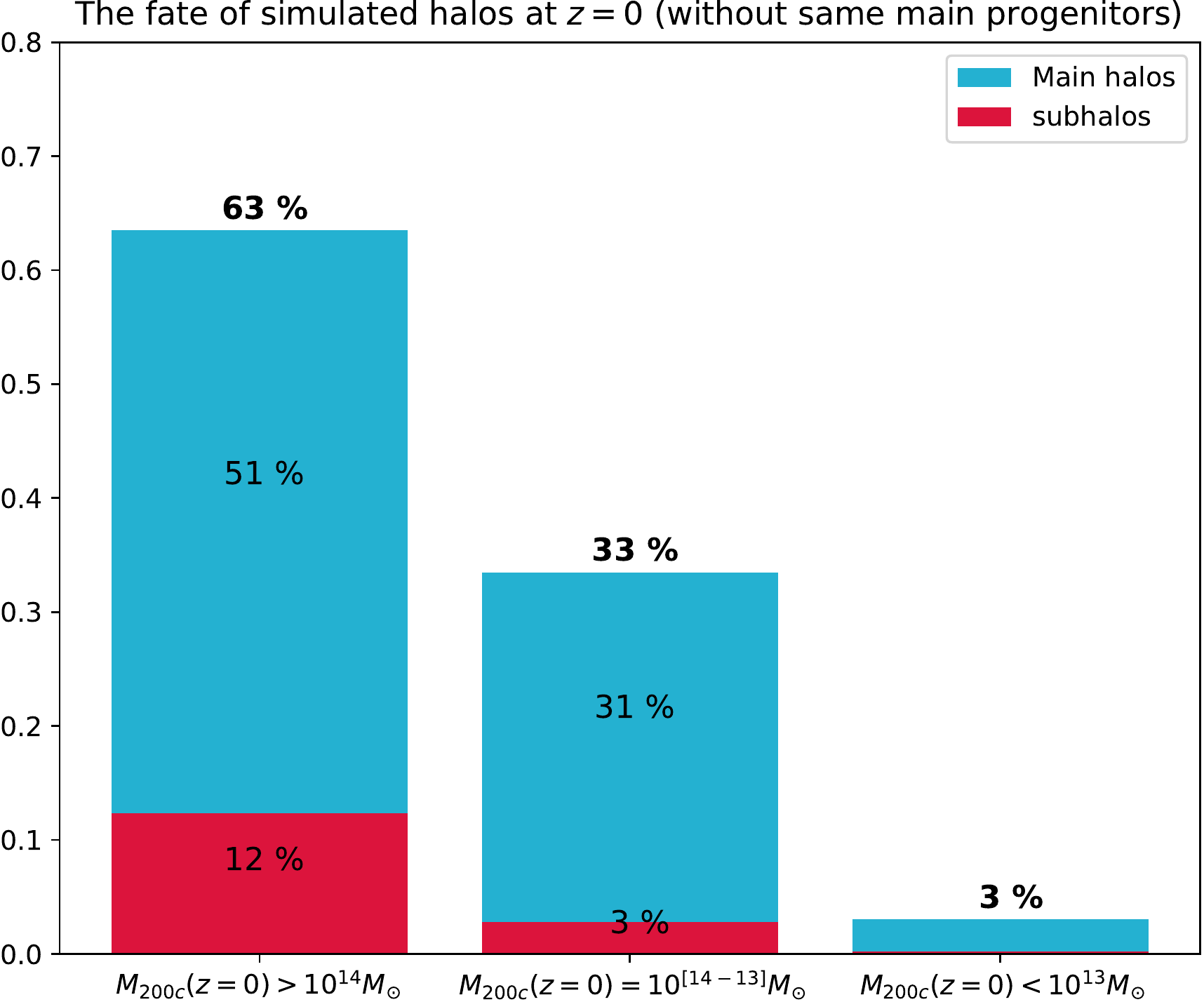}
     \caption{Fractions of 430 distinct (not evolutionary connected) simulated SF halos at high-$z$  that will evolve into clusters with halo mass $M_{200}{>}10^{14} M_{\odot}$, into groups with halo mass $M_{200}=[1-10]\times 10^{13} M_{\odot}$, and into single halos with $M_{200}< 1 \times 10^{13} M_{\odot}$. We
also show the fraction of simulated halos that will become the
main subhalo of a halo at $z$\,=\,0 (turquoise), or a substructure
(subhalo) inside a given halo (red). \label{fig:FATE_without_double}}
\end{figure}

Given that we perform a SFR-based selection of halos at each snapshot from $z=1.3$ to $3$, we might in principle consider the same SF halo at different time step.
It is thus legitimate to wonder whether these \textit{"replicated"} simulated halos affect significantly our statistics on their fate at $z=0$.
To address this issue, we re-perform the analysis described in Sect.~\ref{SEC:FATE}, but after removing all objects that are direct progenitors of the same $z$=0 structure, leaving only the main progenitor the first time (at the highest $z$) it appears.
To find these direct progenitors, we consider the objects which belong to the main progenitor branch of the same merger tree (the main progenitor of each subhalo is defined as the one with the "most massive history" behind it).
For example, in almost each of the 19 snapshot from $z=1.3$ to $3$, we have in our sample the main progenitor of the most massive cluster at $z=0$ (17 times). Through this technique, we find that 173 objects, in our sample of 570 SF halos, are the direct main progenitors of 33 structures at $z=0$. After removing the \textit{"replicated"} halos, the \textit{"cleaned"} halo sample contains 430 distinct high-$z$ SF halos. As shown in Fig. \ref{fig:FATE_without_double},  removing the \textit{"replicated"} halos from our same does not change drastically the results. The fraction of SF halos at high-$z$ that will become galaxy massive clusters at $z=0$ decreases from $72\%$ to $63\%$. As expected most of the \textit{"replicated"} halos are the most massive ones at the observed redshift, that often coincides with the progenitors of most massive structures at $z=0$.

\end{document}